\documentclass[twocolumn]{aastex631}
\usepackage{graphicx}	% Including figure files
\usepackage{amsmath}	% Advanced maths commands
\usepackage{natbib}
\usepackage{multirow} % Table
\usepackage{float} % for figure H command
\usepackage{placeins} % for FloatBarrier
\usepackage{enumerate}
\usepackage{fancyvrb} % fancy verbatim
\usepackage[]{hyperref}
\usepackage{color}
\let\tablenum\relax % deal with the conflict between AASTex and siunitx on tablenum
\usepackage{siunitx}
\usepackage{diagbox}
\usepackage{array}
\usepackage{booktabs}

\newcommand{\asymunc}[3]{\ensuremath{#1_{-#2}^{+#3}}}
\usepackage{xcolor}
\usepackage[normalem]{ulem}
\newcommand{\rem}[1]{
  \bgroup
  \markoverwith{\textcolor{blue}{\rule[0.5ex]{2pt}{0.4pt}}}
  \ULon{#1}
  \egroup
}
\newcolumntype{C}[1]{>{\centering\arraybackslash}m{#1}}
\newcolumntype{M}[1]{>{\centering}m{#1}}

\def\xmm{\textit{XMM-Newton}}
\def\pn{{EPIC-pn}}
\def\om{{OM}}

\def\nustar{\textit{Nustar}}
\def\xspec{\texttt{XSPEC}}
\def\pexrav{\texttt{pexrav}}
\def\zpowerlw{\texttt{zpowerlw}}

\def\zbbody{\texttt{zbbody}}

\def\zcutoffpl{\texttt{zcutoffpl}}

\def\asurv{\texttt{ASURV}}
\def\topcat{\texttt{TOPCAT}}
\def\python{Python}
\defcitealias{Chen+2025a}{paper I}

\begin{document}

\title{A UV to X-Ray View of Soft Excess in Type 1 Active Galactic Nuclei. II. Broadband Correlations}
\shorttitle{Soft X-ray Excess, Hot Corona, and Accretion Disk}
\shortauthors{Chen et al.}

\correspondingauthor{Shi-Jiang Chen \& Jun-Xian Wang}\email{JohnnyCsj666@gmail.com, jxw@ustc.edu.cn}

\author[0009-0004-9950-9807]{Shi-Jiang Chen}
\affiliation{Department of Astronomy, University of Science and Technology of China, Hefei 230026, People's Republic of China}
\affiliation{School of Astronomy and Space Science, University of Science and Technology of China, Hefei 230026, People's Republic of China}

\author[0000-0002-4419-6434]{Jun-Xian Wang}
\affiliation{Department of Astronomy, University of Science and Technology of China, Hefei 230026, People's Republic of China}
\affiliation{School of Astronomy and Space Science, University of Science and Technology of China, Hefei 230026, People's Republic of China}
\affiliation{College of Physics, Guizhou University, Guiyang, Guizhou, 550025, People's Republic of China}

\author[0000-0003-2280-2904]{Jia-Lai Kang}
\affiliation{Department of Astronomy, University of Science and Technology of China, Hefei 230026, People's Republic of China}
\affiliation{School of Astronomy and Space Science, University of Science and Technology of China, Hefei 230026, People's Republic of China}

\author[0000-0003-2573-8100]{Wen-Yong Kang}
\affiliation{Department of Astronomy, University of Science and Technology of China, Hefei 230026, People's Republic of China}
\affiliation{School of Astronomy and Space Science, University of Science and Technology of China, Hefei 230026, People's Republic of China}

\author[0000-0002-9265-2772]{Hao Sou}
\affiliation{Department of Astronomy, University of Science and Technology of China, Hefei 230026, People's Republic of China}
\affiliation{School of Astronomy and Space Science, University of Science and Technology of China, Hefei 230026, People's Republic of China}

\author[0000-0002-2941-6734]{Teng Liu}
\affiliation{Department of Astronomy, University of Science and Technology of China, Hefei 230026, People's Republic of China}
\affiliation{School of Astronomy and Space Science, University of Science and Technology of China, Hefei 230026, People's Republic of China}

\author[0000-0002-4223-2198]{Zhen-Yi Cai}
\affiliation{Department of Astronomy, University of Science and Technology of China, Hefei 230026, People's Republic of China}
\affiliation{School of Astronomy and Space Science, University of Science and Technology of China, Hefei 230026, People's Republic of China}

\author[0000-0001-8515-7338]{Zhen-Bo Su}
\affiliation{Department of Astronomy, University of Science and Technology of China, Hefei 230026, People's Republic of China}
\affiliation{School of Astronomy and Space Science, University of Science and Technology of China, Hefei 230026, People's Republic of China}

% Abstract of the paper
\begin{abstract}
The physical origin of soft X-ray excess (SE) is a long lasting question, with two prevailing theories---``warm corona'' and ``ionized reflection''---dominating the discussion. In the warm corona scenario, SE originates from upscattered disk photons and should therefore correlate strongly with UV emission. Conversely, in the ionized reflection scenario, SE arises from the illumination of the accretion disk by the hot corona and should primarily correlate with the hard X-ray primary continuum (PC). In this second paper of the series, we investigate the correlations among SE, UV and PC, leveraging a sample of 59 unobscured type 1 AGNs compiled in \citet{Chen+2025a}. Our extensive analysis reveals a strong intrinsic correlation between SE and UV that remains robust after controlling for PC ($p_\mathrm{null}\lesssim\num{1e-7}$). In contrast, the correlation between SE and PC is weaker but still statistically significant ($p_\mathrm{null}\lesssim\num{5e-2}$). These findings suggest that, in addition to ionized reflection---a natural outcome of the hot corona illuminating the disk---a warm corona component is essential, and may even dominate, in producing the soft excess. Additionally, we report a mild anti-correlation between SE strength ($q$) and PC photon index ($\Gamma_\mathrm{PC}$) ($p_\mathrm{null}=\num{1e-2}$), suggesting a potential competition between the warm and hot coronae. Finally, we find that the $\Gamma_\mathrm{PC}$ values we derived with SE properly incorporated exhibit a much weaker correlation with $\lambda_\mathrm{Edd}$ ($p_\mathrm{null}=\num{2e-2}$) than previously reported in the literature. This highlights the critical role of accurately modeling SE in studies of the $\Gamma_\mathrm{PC}$--$\lambda_\mathrm{Edd}$ relation.
\end{abstract}


\section{Introduction}\label{sec:Intro}
The soft X-ray excess (SE) is a spectral feature commonly observed in type 1 AGNs \citep[e.g.,][]{Pravdo+1981,Singh+1985,Arnaud+1985,George+2000,Reeves&Turner2000,Mineo+2000,Porquet+2004}. Despite its ubiquity \citep[e.g.,][]{Bianchi+2009}, the physical origin of SE remains unclear. Two prevailing yet potentially competing theories dominate our current understanding of SE \citep[e.g.,][]{Kara&Garcia2025}: the ``warm corona'' model \citep[e.g.,][]{Magdziarz+1998,Mehdipour+2011,Rozanska2015,Gronkiewicz&Rozanska2020,Palit+2024,Kawanaka&Mineshige2024} and the ``ionized disk reflection'' model \citep[e.g.,][]{Ross&Fabian1993,Ballantyne+2001,Miniutti&Fabian2004,Ross&Fabian2005,Crummy+2005,Merloni+2006,Garcia&Kallman2010,Dovciak+2011,Bambi+2021}. In the first scenario, disk UV photons are upscattered into the soft X-ray band via a warm corona that either overlays the standard disk \citep[e.g.,][]{Petrucci+2013,Petrucci+2018} or replaces it within a specific radius \citep[e.g.,][]{Done+2012,Kubota&Done2018,Hagen+2024}. In contrast, the ionized disk reflection model posits that SE arises from relativistically blurred soft X-ray emission lines generated when the primary hard X-ray continuum (PC) from the corona illuminate the highly ionized inner regions of the accretion disk.

% so far, model comparison is a general approach ... but not enough
The most fundamental approach to identifying the nature of SE involves fitting physical models to observed spectra and comparing their outcomes \citep[e.g.,][]{Dewangan+2007,Fabian+2012,Chiang+2015,Jiang+2018,Tripathi+2019,Middei+2020,Liu+2020,Xu+2021a,Waddell+2023}. However, current X-ray spectral quality often permits both warm corona and ionized reflection models to fit data equivalently well \citep[e.g.,][]{Jin+2016,Garcia+2019,Xu+2021,Chalise+2022}. With ongoing advancements in physically sophisticated \citep[e.g.,][]{Garcia+2014,Ding+2024} and flexible \citep[e.g.,][]{Garcia+2016,Jiang+2019} models, distinguishing between these scenarios via spectral analysis alone may become increasingly challenging, at least until next-generation X-ray instruments with higher spectral resolution and greater sensitivity (e.g., New Athena, \citealt{Cruise+2024}; eXTP, \citealt{Zhang+2025a,Zhou+2025}). 

% an alternative way: lum corr
A promising alternative approach involves examining broadband SE luminosity correlations. Different physical mechanisms for SE predict distinct luminosity correlation patterns. 
In the ionized disk reflection scenario, where SE results from hot corona illumination, the SE luminosity should correlate more strongly with the primary continuum. Conversely, in the warm corona scenario, where SE originates from upscattered disk photons, SE luminosity is thus expected to correlate more strongly with UV emission (although a certain degree of SE--PC correlation might still arise due to the coupling between the warm and hot coronae, e.g., \citealt{Noda&Done2018}). By measuring luminosities without relying on specific spectral models, this method offers an intuitive and model-independent means to probe the nature of SE. This approach has been applied to several individual sources with time-domain data \citep[e.g.,][]{Mehdipour+2011,Ursini+2020a,Mehdipour+2023,Partington+2024} to investigate how SE variability correlates with changes in UV and PC emission. These studies generally support the dominance of the warm corona scenario. However, they are limited to individual cases and do not explore the systematic variation of SE across different sources. A comprehensive analysis of a large sample, examining broadband luminosity correlations among diverse AGNs, remains lacking. Such an investigation would provide critical insights into the global properties of SE and help identify the dominant mechanisms driving SE across the AGN population. Furthermore, correlating SE strength with other physical parameters (e.g., Eddington ratio, PC photon index) offers a valuable avenue for exploring interactions between SE, the accretion disk, and the hot corona \citep[e.g.,][]{Walter&Fink1993,Liu&Qiao2010,Boissay+2016,Gliozzi&Williams2020,Waddell&Gallo2020,Nandi+2023}.

In the first paper of this series (\citealt{Chen+2025a}, hereafter \citetalias{Chen+2025a}) we compiled a sample of 59 type 1 AGNs from \xmm~archive. The X-ray spectra of these sources show neither strong absorption (neutral, warm, or partial) nor significant soft X-ray emission lines, making them ideal for studying the intrinsic properties of SE. In \citetalias{Chen+2025a}, we discovered that the spectral profile of the SE is correlated with its strength relative to PC. Stronger SE exhibit more extended, power-law-like spectral profiles, while weaker SE have profiles more akin to blackbody emission. This finding indicates that the SE cannot be explained solely by disk reflection, which produces mostly blackbody-like SE. Instead, it supports a hybrid model in which both the warm corona and reflection contribute to the SE.

The sample of \citetalias{Chen+2025a} was selected to have simultaneous \xmm~X-ray and UV observations, thus enabling comprehensive broadband analyses. Building on this foundation, in this second paper of the series we investigate how SE luminosity and strength correlate with broadband SED parameters. In \S\ref{sec:SampDR} we briefly review our sample selection and spectral models. To provide an intuitive illustration of the connection among SE, UV, and PC radiation, we present the UV-to-X-ray SEDs of the sample in \S\ref{sec:SED}. Subsequently, in \S\ref{sec:2Dlum} and \S\ref{sec:3Dlum}, we quantitatively examine the luminosity correlations among SE, UV and PC to probe the nature of soft excess. Finally in \S\ref{sec:Phys} and \S\ref{sec:GammaLambda} we explore correlations among SE strength and key physical parameters, such as the Eddington ratio and PC photon index, and discuss their implications.

Following the convention in \citetalias{Chen+2025a}, we abbreviate the hard X-ray primary continuum as ``primary continuum'' or ``PC'', and the soft X-ray excess as ``soft excess'' or ``SE''. Unless otherwise noted, these two luminosities (without additional subscripts) refer to the quantities derived from model 3 (see \S\ref{sec:SampDR}). The UV luminosity refers to luminosity at \qty{2500}{\AA} (\unit{erg.s^{-1}}), converted from \om~UVW1 flux. Throughout this paper, we assume a cosmology with $H_0=\qty{70}{km.s^{-1}.Mpc^{-1}}$, $\Omega_\mathrm{m}=0.3$ and $\Omega_\Lambda=0.7$.

\section{Overview of the sample and spectral models} \label{sec:SampDR}
The sample selected in \citetalias{Chen+2025a} consists of 59 local ($z<0.4$) unobscured type 1 AGNs with simultaneous \xmm~\pn~and \om~(UVW1) observation. Careful consideration was given to exclude sources with strong neutral, warm, or partial absorption, ensuring a reliable assessment of the intrinsic SE. The contribution of the host galaxy to the observed UVW1 band is negligible for all sources. Additionally, the sample includes only bright AGNs, ensuring sufficient data quality for robust spectral analysis. Each \pn~spectrum was fitted using three distinct phenomenological models:

\begin{verbatim}
1: phabs*zphabs*(zbbody+pexrav+zgauss1+zgauss2) 
    
2: phabs*zphabs*(zpowerlw+pexrav+zgauss1+zgauss2)

3: phabs*zphabs*(zcutoffpl+pexrav+zgauss1+zgauss2)
\end{verbatim}

The model \pexrav~\citep{Magdziarz&Zdziarski1995} is adopted for the X-ray primary continuum, plus the neutral reflection component. The iron line is represented by two Gaussian components: one for the narrow core, with the centroid energy fixed at \qty{6.4}{keV} and the line width fixed at \qty{0.019}{keV} \citep[e.g.,][]{Shu+2010,Kang2020}; and another for the broad component, with the centroid energy allowed to vary within \qtyrange[range-phrase=--,range-units=single]{5}{6.5}{keV} and the line width constrained to $<\qty{1}{keV}$.

To characterize the SE, whose physical nature remains uncertain, \citetalias{Chen+2025a} employed three distinct phenomenological models: blackbody (\zbbody, model 1), power-law (\zpowerlw, model 2), or cut-off power-law (\zcutoffpl, model 3). The analysis revealed that $29\%$ of the sources exhibit SE with blackbody-like profiles, while $71\%$ favor power-law-like profiles. The cut-off power-law model, which mimics a blackbody at low cut-off energies and transitions to a power-law at higher cut-off energies, provided the most unbiased and overall best fits.

Using the best-fit parameters from models 1, 2, and 3, the following quantities were derived: SE luminosity ($L_\mathrm{SE,bb}$, $L_\mathrm{SE,po}$, $L_\mathrm{SE,cpl}$), PC luminosity ($L_\mathrm{PC,bb}$, $L_\mathrm{PC,po}$, $L_\mathrm{PC,cpl}$), and PC photon index ($\Gamma_\mathrm{PC,bb}$, $\Gamma_\mathrm{PC,po}$, $\Gamma_\mathrm{PC,cpl}$). The UVW1 flux was converted into UV luminosity at \qty{2500}{\AA} ($L_\mathrm{UV}$) using a K-correction with a UV spectral slope of $\alpha=\num{0.65}$ ($F_\nu\sim\nu^{-\alpha}$; \citealt{Natali+1998}). The BH masses ($M_\mathrm{BH}$) for the sources were compiled from the literature. Combining $M_\mathrm{BH}$ and $L_\mathrm{UV}$,  the Eddington ratio ($\lambda_\mathrm{Edd}$) was estimated, assuming a UV-to-bolometric correction factor of \num{2.75} \citep{Krawczyk+2013}. A full catalog and details are provided in the appendix of  \citetalias{Chen+2025a}. We note that $\lambda_\mathrm{Edd}$ may fluctuate around the true Eddington ratio ($\lambda_{\mathrm{Edd},true}$), since the adopted correction factor of \num{2.75} is a sample average and individual sources likely scatter around it.

Our UV-based $\lambda_\mathrm{Edd}$, as defined above, is 2.75 times the Eddington-scaled UV luminosity. The $\lambda_\mathrm{Edd}$ for our sample could also be estimated from the X-ray luminosity. However, because the X-ray bolometric correction factor $\kappa_\mathrm{X}$ shows large dispersion and only a weak dependence on X-ray luminosity \citep[e.g.,][]{Vasudevan&Fabian2007}, we follow \citealt{Nandi+2023} and directly adopt the Eddington-scaled primary continuum luminosity ($L_\mathrm{PC,0.5-10}/L_\mathrm{Edd}$) as an alternative proxy for the Eddington ratio (see the relevant results presented in \S\ref{sec:Phys}, \S\ref{sec:GammaLambda}, and Appendix \ref{sec:Edd_X}).

\section{UV-to-X-ray SEDs}\label{sec:SED}
%%%%%%%%%%%%%% The Figures %%%%%%%%%%%%%%
\begin{figure*}[htbp]
    \centering
\includegraphics[width=2.0\columnwidth]{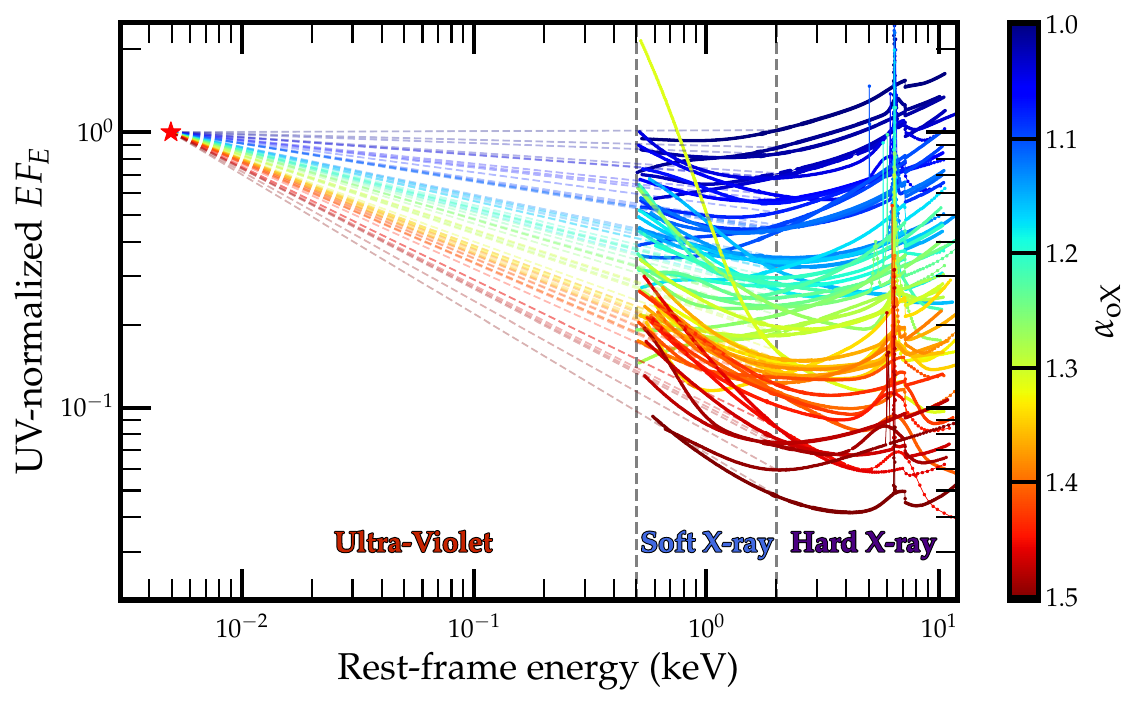}
    \caption{Normalized UV-to-X-ray SEDs for our sample. The absorption corrected best-fit X-ray model spectrum (model 3) is shifted to the rest frame, scaled to the simultaneous rest frame \qty{2500}{\AA} flux, and color coded based on $\alpha_\mathrm{oX}$ (the power-law slope of the dashed color lines).}
    \label{fig:UV_norm_spec}
\end{figure*}
\begin{figure}
    \includegraphics[width=1.0\columnwidth]{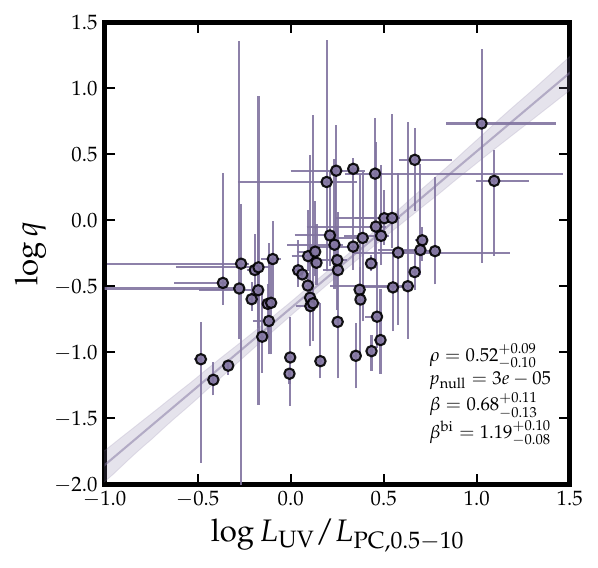}
    \caption{Relation between SE strength ($q$) and UV--PC luminosity ratio ($L_\mathrm{UV}/L_\mathrm{PC,0.5-10}$). Shown are the Spearman’s $\rho$, null hypothesis probability $p_\mathrm{null}$, OLS slope $\beta$, and OLS bisector slope $\beta^\mathrm{bi}$. The shaded region marks the $1\sigma$ confidence interval of the bisector fit. Unless noted otherwise, the same legend format applies to all scatter plots in this paper.}
    \label{fig:q_UVPC}
\end{figure}
\begin{figure}
    \includegraphics[width=1.0\columnwidth]{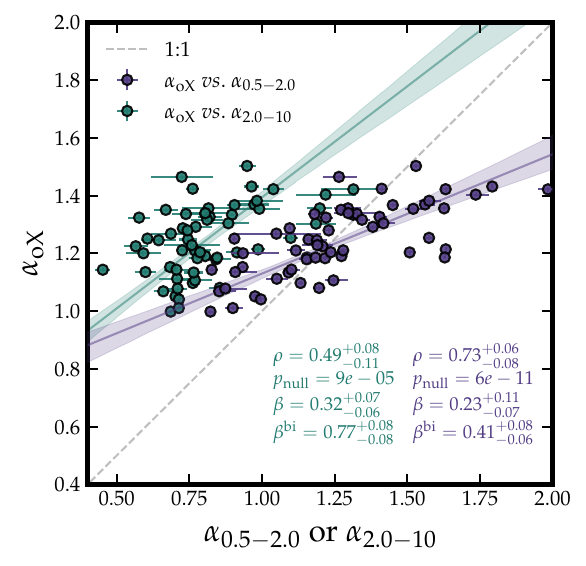}
    \caption{Relation between $\alpha_\mathrm{oX}$ and soft X-ray spectral slope ($\alpha_\mathrm{0.5-2}$, purple) or hard X-ray spectral slope ($\alpha_\mathrm{2-10}$, green). A 1:1 reference line is shown in gray for comparison.}
    \label{fig:alphaX}
\end{figure}
%%%%%%%%%%%%%%%%%%%%%%%%%%%%%%%%%%%%%%%%%
Fig. \ref{fig:UV_norm_spec} present the UV-to-X-ray SEDs for our sample, normalized at \qty{2500}{\AA}, and color coded according to $\alpha_\mathrm{oX}$, a parameter quantifying the relative strength of the disk to corona, calculated as $-0.3838\times\log \left(F_\text{2keV}/F_\text{2500\AA}\right)$ \citep[e.g.,][]{Sobolewska+2009a}. Such a plot provides a direct visualization of the connection between the SE and the broadband SED.

A key takeaway from Fig. \ref{fig:UV_norm_spec} is that sources with stronger UV relative to hard X-rays tend to exhibit more pronounced SE. By ``more pronounced SE'' we mean not only stronger SE relative to the underlying PC, but also steeper soft X-ray spectral slope. These observations suggest that the SE may be more closely tied to UV emission than to hard X-rays. However, further quantitative correlation analyses are necessary to uncover the true nature of SE, which will be explored in \S\ref{sec:correlation}.

\section{Correlation analysis and implications}\label{sec:correlation}
In this section, we analyze correlations in our sample to investigate the nature of SE. We use Spearman’s rank coefficient ($\rho$) for its robustness to outliers, with $1\sigma$ errors estimated via bootstrapping. Each scatter plot shows $\rho$, the null-hypothesis probability $p_\mathrm{null}$, the Ordinary Least Squares (OLS) slope $\beta$, and the OLS bisector slope $\beta^\mathrm{bi}$ \citep[e.g.,][]{Isobe&Feigelson1990}. The $1\sigma$ confidence range of the bisector fit, derived from bootstrapping, is shown as a shaded region.

We begin by quantifying the trends seen in Fig. \ref{fig:UV_norm_spec} using two approaches. First, we compare $q$---defined as luminosity ratio of SE to PC in \qtyrange[range-phrase=--,range-units=single]{0.5}{2}{keV} (see \S\ref{sec:Phys})---and $L_\mathrm{UV}/L_\mathrm{PC,0.5-10}$, which characterizes the UV to hard X-ray ratio (Fig. \ref{fig:q_UVPC}). Both SE and PC luminosities are derived from model 3. We find a strong correlation, $\rho=\asymunc{0.52}{0.10}{0.09}$, indicating SE becomes more pronounced as UV strengthens relative to the hard X-ray.

Second, motivated by the observation in Fig. \ref{fig:UV_norm_spec} that the soft and hard X-ray slope show different trends with the UV-to-X-ray slope, we measure these slopes in a fully model-independent manner---that is, without invoking any spectral decomposition---and compare them in Fig. \ref{fig:alphaX}. Here, the soft X-ray slope, $\alpha_\mathrm{0.5-2}$, is derived by fitting the observer-frame \qtyrange[range-phrase=--,range-units=single]{0.5}{2}{keV} spectrum with the model \texttt{phabs*zphabs*zpowerlw}, fixing the Galactic absorption (\texttt{phabs}) to $N_\mathrm{H,Gal}$ and allowing the intrinsic absorption (\texttt{zphabs}) to vary. The hard X-ray slope, $\alpha_\mathrm{2-10}$, is obtained by fitting \texttt{phabs*zpowerlw} to the \qtyrange[range-phrase=--,range-units=single]{2}{10}{keV} range.

In Fig. \ref{fig:alphaX}, we see tight correlation between the soft X-ray spectral slope $\alpha_\mathrm{0.5-2}$ and $\alpha_\mathrm{oX}$, in agreement with previous studies \citep{Walter&Fink1993,Liu&Qiao2010,Grupe+2010}. Additionally, many sources have $\alpha_\mathrm{0.5-2}$ values very close to $\alpha_\mathrm{oX}$, consistent with the pattern seen in Fig. \ref{fig:UV_norm_spec} where the soft X-ray slope aligns closely the UV-to-X-ray connecting line (dashed line), although the slope of their tight correlation is different from 1:1. Meanwhile, the hard X-ray spectral slope $\alpha_\mathrm{2-10}$ is systematically harder than $\alpha_\mathrm{oX}$, and exhibits a weaker correlation.

These preliminary quantitative analyses already provide useful insights, reinforcing the relationship illustrated in Fig. \ref{fig:UV_norm_spec}, and suggesting that the SE--UV association is stronger than that between SE and PC. To more rigorously quantify this, we will conduct direct correlation analyses among SE, UV and PC luminosities in \S\ref{sec:2Dlum}. We characterize SE by its luminosity in \qtyrange[range-phrase=--,range-units=single]{0.5}{2}{keV}. The UV luminosity at \qty{2500}{\AA} serves as a proxy for accretion disk. The hot corona, which is responsible for the PC emission, is characterized by the \pexrav~luminosity in \qtyrange[range-phrase=--,range-units=single]{2}{10}{keV}, which primarily represents the PC emission and is minimally affected by degeneracies with SE. The neutral reflection component contributes $\lesssim 20\%$ to the total \pexrav~luminosity in this range, and excluding it does not change our results.

\subsection{Bivariate correlation analysis
%The 2D luminosity correlations
}\label{sec:2Dlum}
%%%%%%%%%%%%%% The Figures %%%%%%%%%%%%%%
\begin{figure*}
    \centering
    \includegraphics[width=2.0\columnwidth]{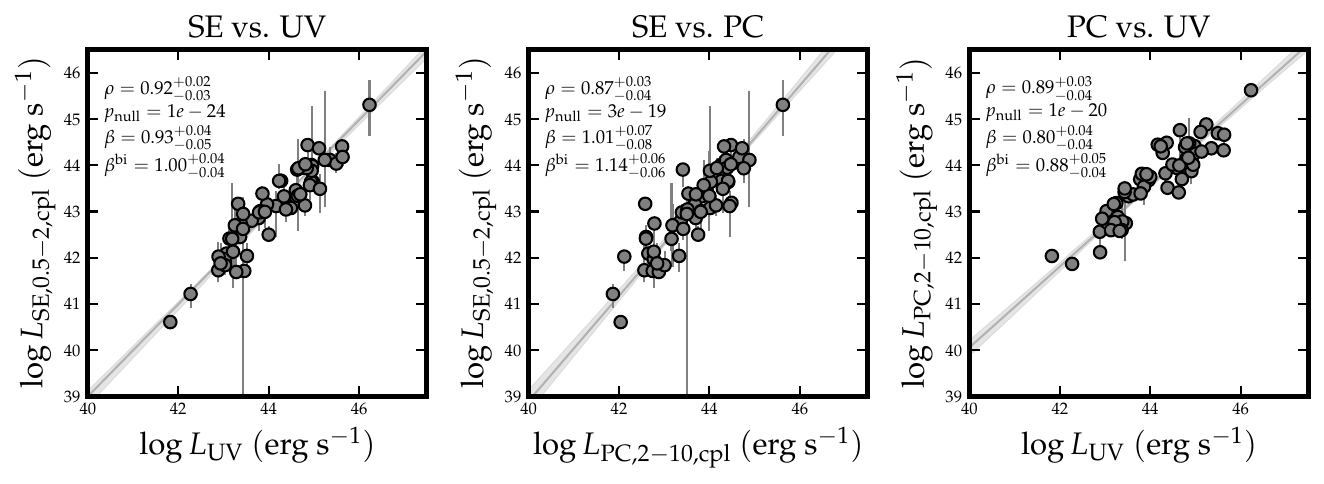}
    \caption{Luminosity-luminosity correlation for SE vs. UV (left), SE vs. PC (middle), and PC vs. UV (right). Model 3 is used to derive SE and PC luminosities.}
    \label{fig:lum}
\end{figure*}
%%%%%%%%%%%%%%%%%%%%%%%%%%%%%%%%%%%%%%%%%
We begin by investigating the bivariate correlations between the luminosities of the three key components: SE vs. UV, SE vs. PC, and PC vs. UV. Note while many previous studies have explored the luminosity correlation between soft X-ray emission and UV luminosity \citep[e.g.,][]{Steffen+2006,Young+2010,Lusso+2010,Risaliti&Lusso2015,Nour&Sriram2022}, few works have specifically examined the correlation between the soft X-ray ``excess'' (i.e., the residual component after subtracting the underlying PC in the soft X-ray) and UV luminosity. Moreover, systematic comparisons involving PC remain sparse.

From left to right in Fig. \ref{fig:lum}, we present the correlations between SE and UV, SE and PC, and PC and UV luminosities. The results shown are based on model 3, as no significant differences were observed with the other models. All three panels reveal very strong correlations, with Spearman’s rank correlation coefficients $\rho$ exceeding \num{0.87} and $p_\mathrm{null}$ below \num{3e-19}. We do not observe statistically significant difference in Spearman's $\rho$ among the three correlations, suggesting strong coupling among all three components. Examining the bisector slopes, we find that the SE--UV correlation follows a nearly 1:1 relation ($\beta^\mathrm{bi}=\asymunc{1.00}{0.04}{0.04}$). In contrast, the bisector slopes of the SE--PC and PC--UV correlations deviate from unity, with $\beta^\mathrm{bi}=\asymunc{1.14}{0.06}{0.06}$,  $\asymunc{0.88}{0.04}{0.05}$, respectively.

The nearly 1:1 relationship between SE and UV luminosities has interesting implications. It suggests that the SE luminosity scales linearly with UV luminosity across AGNs of varying brightness, implying a physical connection between the SE component and the accretion disk. Combined with recent findings \citep{Cai&Wang2023,Cai2024}  demonstrating that the optical--NUV--EUV (disk) SED shape of quasars remains remarkably consistent across different luminosities, this result may extend the observed constancy to the soft X-ray regime and for local AGNs. In comparison, the bisector slope between PC and UV luminosities is significantly less than 1. This is consistent with the well-established trend that more luminous AGNs tend to exhibit weaker X-ray emission relative to their bolometric luminosity \citep[e.g.,][]{Vasudevan&Fabian2009,Netzer2019,Duras+2020,Ballantyne+2024}, suggesting a non-linear dependence between the two emissions.

\subsection{Statistical decomposition of SE--UV--PC luminosity coupling}\label{sec:3Dlum}
Despite their utility, the bivariate luminosity correlations discussed in \S\ref{sec:2Dlum} have limitations. First, they may be influenced by redshift effects. Second, since the accretion system comprises three interrelated components (SE, UV, and PC), analyzing the correlations between only two of them inherently neglects the role of the third. To address these issues, we investigate the intrinsic SE--UV and SE--PC correlations using three distinct methods, as illustrated below.

\subsubsection{Correlations between normalized luminosities}\label{sec:3Dlum_ratio}
%%%%%%%%%%%%%% The Figures %%%%%%%%%%%%%%
\begin{figure*}
    \includegraphics[width=2.0\columnwidth]{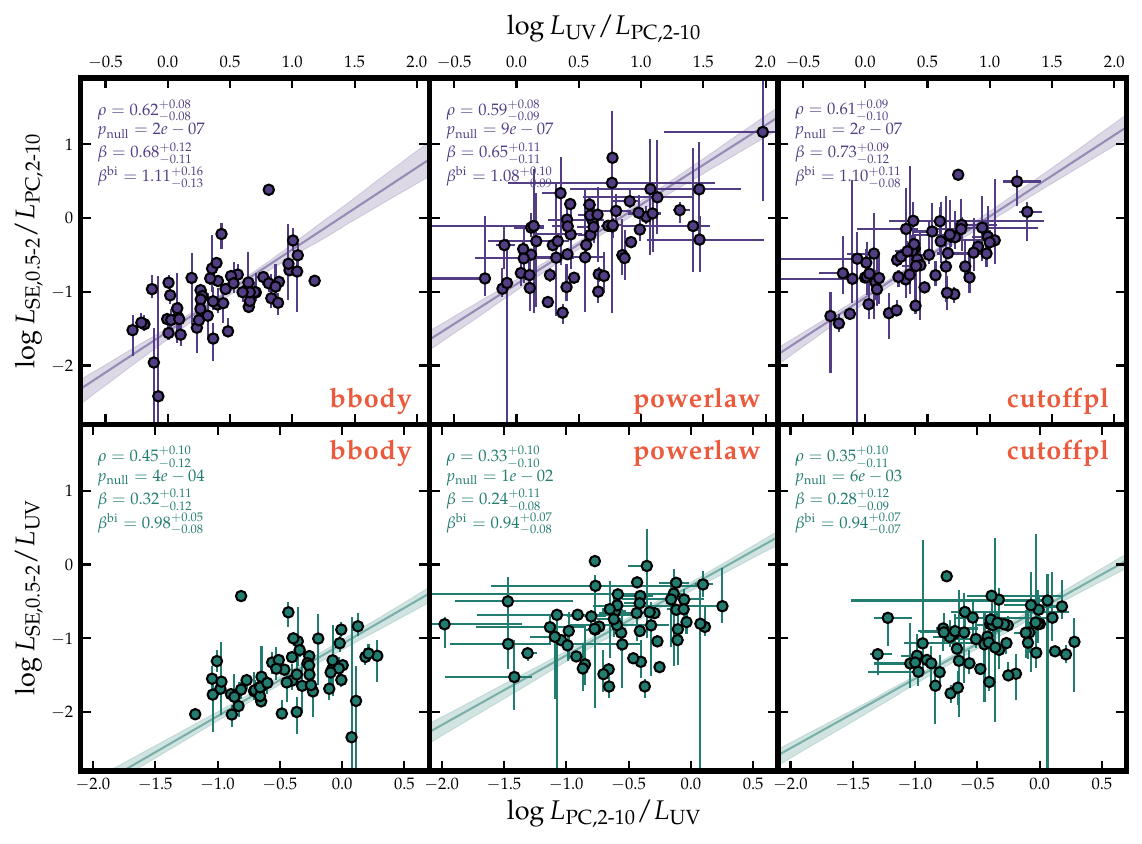}
    \caption{Intrinsic SE--UV and SE--PC relations as revealed by luminosity-ratio correlations. The top three panels (in purple) illustrate the intrinsic correlation between SE and UV luminosities, with both scaled by PC luminosity, under different SE models (blackbody, power-law, and cut-off power-law). The bottom three panels (in green) similarly show the intrinsic correlation between SE and PC luminosities, scaled by UV luminosity. The results highlight a strong intrinsic link between SE and UV luminosities ($p_\mathrm{null}\sim\num{2e-7}$), even after removing PC contributions. The correlation between SE and PC is also statistically significant ($p_\mathrm{null}\sim\num{6e-3}$) but is generally weaker than the SE--UV correlation.}
    \label{fig:lumratio}
\end{figure*}
%%%%%%%%%%%%%%%%%%%%%%%%%%%%%%%%%%%%%%%%%
The SE luminosity may depend intrinsically on both UV and PC luminosities. To 
isolate these effects, we normalize SE and UV by PC and examine their relation ($\log L_\mathrm{SE,0.5-2}/L_\mathrm{PC,2-10}$ vs. $\log L_\mathrm{UV}/L_\mathrm{PC,2-10}$), which provides a first-order view of how SE responds to UV at fixed PC. Conversely, to probe the intrinsic SE--PC link independent of UV, we analyze $\log L_\mathrm{SE,0.5-2}/L_\mathrm{UV}$ vs. $\log L_\mathrm{PC,2-10}/L_\mathrm{UV}$. This luminosity ratio method provides a simple yet effective way to visualize the intrinsic interplay among SE, UV, and PC.

The results are shown in Fig. \ref{fig:lumratio}. The top three panels (purple) illustrate the intrinsic SE--UV correlations for three SE models (blackbody, power-law, and cut-off power-law), while the bottom three panels (green) depict the intrinsic SE--PC correlations. The intrinsic SE--UV correlation remains strong, with Spearman's $\rho$ exceeding \num{0.59} ($p_\mathrm{null}<\num{9e-7}$), even after removing PC contributions. In contrast, while the intrinsic correlation between SE and PC also persists, it is noticeably weaker than that between SE and UV. These conclusions are robust across all three SE models.

\subsubsection{Partial correlation}\label{sec:3Dlum_parcorr}
%%%%%%%%%%%%%% The Figures %%%%%%%%%%%%%%
\begin{figure}
    \includegraphics[width=1.0\columnwidth]{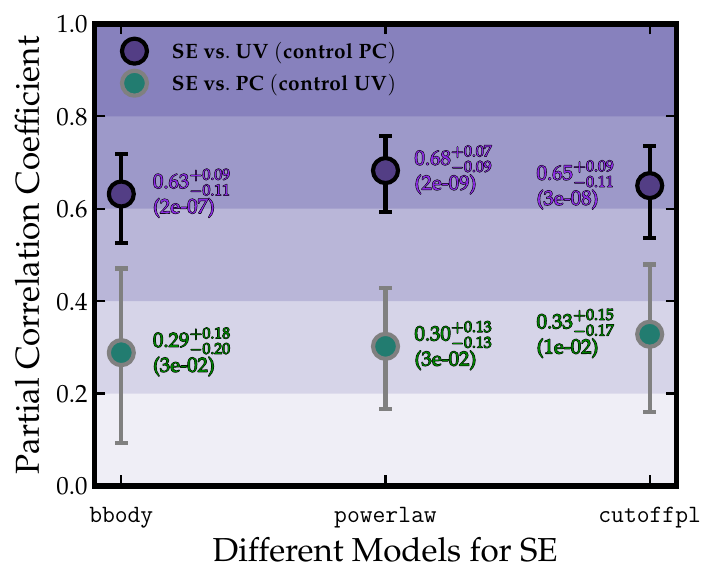}
    \caption{Visualization of partial correlation analysis. The plot displays the partial correlation coefficients (Spearman's $\rho$, Y-axis) and their corresponding p-values, for all three SE models (X-axis).
    %: blackbody, power-law, and cut-off power-law). 
    The purple points represent the partial correlation between SE and UV luminosities after controlling for PC luminosity, while the green points indicate the partial correlation between SE and PC luminosities after controlling for UV luminosity. Across all SE models, the partial correlation between SE and UV (purple) is consistently stronger than the correlation between SE and PC (green).}
    \label{fig:parcorr}
\end{figure}
%%%%%%%%%%%%%%%%%%%%%%%%%%%%%%%%%%%%%%%%%
Another approach to deciphering the intrinsic correlations among three variables is through partial correlation analysis. Mathematically, the partial correlation between variables $X$ and $Y$, while controlling for a third variable $Z$, is given by\footnote{\url{https://en.wikipedia.org/wiki/Partial_correlation\#Using_recursive_formula}}:
\begin{equation}
    \rho_{XY\cdot Z}=\frac{\rho_{XY}-\rho_{XZ}\rho_{ZY}}{\sqrt{1-\rho_{XZ}^2}\sqrt{1-\rho_{ZY}^2}}
\end{equation}
where $\rho_{XY}$ is the Spearman correlation coefficient between $X$ and $Y$; and similar for $\rho_{XZ},\ \rho_{YZ}$.

We apply this method to the variables $\log L_\mathrm{SE,0.5-2}$, $\log L_\mathrm{PC,2-10}$, and $\log L_\mathrm{UV}$, and present the results in Fig. \ref{fig:parcorr}. Since both $\log L_\mathrm{SE,0.5-2}$ and $\log L_\mathrm{PC,2-10}$ depend on the SE model, we use all three models introduced in \S\ref{sec:SampDR}, as labeled along the X-axis. The partial correlation between SE and UV, controlling for PC, is shown as purple filled circles, while that between SE and PC, controlling for UV, is shown as green filled circles. Associated p-values\footnote{\url{https://online.stat.psu.edu/stat505/lesson/6/6.3}} are also provided in parentheses.

The findings from the partial correlation analysis are consistent with those derived in \S\ref{sec:3Dlum_ratio}, despite their different mathematical frameworks. We observe a strong partial correlation ($\rho>0.63$, $p_\mathrm{null}<\num{2e-7}$) between SE and UV, after controlling for PC. This indicates that, for two AGNs with similar PC luminosities, the one with brighter UV tends to exhibit stronger SE. In contrast, the partial correlation between SE and PC is relatively weaker, consistently falling below that of SE and UV across all three SE models. These results suggest that UV luminosity is more influential than PC in producing SE.

\subsubsection{Multiple linear regression}
%%%%%%%%%%%%%% The Figures %%%%%%%%%%%%%%
\begin{figure}
    \includegraphics[width=1.0\columnwidth]{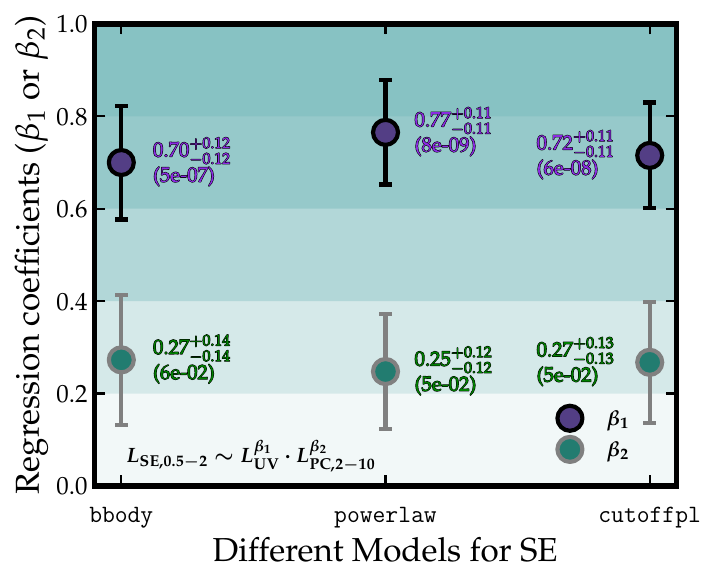}
    \caption{Multiple linear regression slopes of SE on UV (purple) and PC (green) luminosities. Similar to Fig. \ref{fig:parcorr}, but with the Y-axis showing regression slopes from the relation $L_\mathrm{SE,0.5-2}\sim L_\mathrm{UV}^{\beta_1}\cdot L_\mathrm{PC,2-10}^{\beta_2}$. SE depends more strongly on UV ($\beta_1\sim0.72$) than on PC ($\beta_2\sim0.27$).}
    \label{fig:multireg}
\end{figure}
%%%%%%%%%%%%%%%%%%%%%%%%%%%%%%%%%%%%%%%%%
To further quantify the dependencies of SE on PC and UV, we perform a multilinear regression, employing the \python~module \verb|scikit-learn.linear_model| \citep{Pedregosa+2011}. The regression assumes the following functional form: 
\begin{equation}
    \log L_\mathrm{SE,0.5-2}=\beta_1\times\log L_\mathrm{UV}+\beta_2\times \log L_\mathrm{PC,2-10}+\mathrm{const}
\end{equation}
or equivalently:
\begin{equation}
    L_\mathrm{SE,0.5-2}\sim L_\mathrm{UV}^{\beta_1}\cdot L_\mathrm{PC,2-10}^{\beta_2}
\end{equation}
The values of $\beta_1$ and $\beta_2$ are displayed in Fig. \ref{fig:multireg}, with the corresponding p-values indicated in parentheses. Consistent with the results in Fig. \ref{fig:parcorr}, SE exhibits a much stronger dependence on UV ($\beta_1 \gtrsim 0.70$) compared to PC ($\beta_2 \sim 0.27$).

\subsubsection{Implications on the nature of SE}\label{sec:3Dlum_impl}
The physical nature of SE remains a subject of active debate. One intuitive explanation is the ``ionized disk reflection'' model. In this scenario, a fraction of hard X-ray photons emitted by the hot corona illuminates the disk. As long as the disk surface is ionized, the reflected spectrum in the soft X-ray band becomes rich in emission lines, which are then smoothed by relativistic effects, producing an apparent excess. Such an interpretation is supported by the detection of ``soft X-ray lag'' in many sources \citep[e.g.][]{Emmanoulopoulos+2011,Kara+2013,Zoghbi+2015,Wilkins+2021,Bambi+2021,Hancock+2022}, where the variation of soft X-ray lags behind that of hard X-ray, attributed to light-crossing time. The weak yet non-negligible intrinsic correlation of SE and PC as revealed in our sample (Fig. \ref{fig:parcorr}) also supports the presence of such a reflection component.

However, a critical question of ionized reflection is whether it alone can account for the entire observed SE. As argued by several authors (\citealt{Dewangan+2007,Porquet+2018,Chen+2025b,Mallick+2025}, and our \citetalias{Chen+2025a}), an extra component is needed. For instance, our simulations in \citetalias{Chen+2025a} showed that the observed parameter space of SE strength and shape cannot be reproduced by ionized reflection alone, while \citet{Mallick+2025} found that the reflection component constrained from hard X-ray underpredicts the observed SE in some sources. Furthermore, the strong intrinsic SE--UV correlation, compared to the weaker SE--PC correlation---evident from normalized luminosity correlations (Fig. \ref{fig:lumratio}), partial correlation analysis (Fig. \ref{fig:parcorr}), and multiple linear regression (Fig. \ref{fig:multireg})---indicates that in addition to a reflection component, the upscattered disk photons through warm corona are essential, and likely dominant, in producing the observed SE.

Therefore, our analysis in this work provides direct observational support for a hybrid scenario, in which the SE is primarily produced by a warm corona, with a secondary contribution from the ionized reflection. This interpretation also aligns well with recent theoretical studies highlighting the potential coexistence of these components \citep[e.g.,][]{Petrucci+2020,Ballantyne2020,Ballantyne&Xiang2020,Xiang+2022}, proposed to overcome the limitations of single-component interpretations \citep[e.g.,][]{Boissay+2014,Garcia+2019,Liu+2020,Xu+2021,Mallick+2025,Porquet+2025}.

We note that while we interpret the intrinsic SE--PC correlation as indicative of ionized reflection, it does not fully rule out a pure warm corona origin. For instance, soft X-ray photons produced by a warm corona could partially feed the hot corona, or even both SE and PC may arise from the same Comptonization structure, with SE produced by fewer scatterings \citep[e.g.,][]{Nandi+2021}. Future theoretical work is needed to test whether either scenarios alone (without reflection) is sufficient to reproduce the observed SE--PC correlation in Fig. \ref{fig:parcorr}.

\subsection{SE strength vs. physical parameters}\label{sec:Phys}
To gain deeper insights into the nature of SE and its interplay with the accretion system, we examine its relation with key physical parameters, specifically the (UV-based) Eddington ratio ($\lambda_\mathrm{Edd}$) and the PC photon index ($\Gamma_\mathrm{PC}$). Following previous studies \citep[e.g.,][]{Boissay+2016,Gliozzi&Williams2020,Ballantyne+2024}, we quantify SE strength using $q$, defined as the ratio between SE and the underlying PC in the \qtyrange[range-phrase=--,range-units=single]{0.5}{2}{keV} range, i.e., $q\equiv\frac{L_\mathrm{SE,0.5-2}}{L_\mathrm{PC,0.5-2}}$.

\subsubsection{$q$ vs. $\lambda_\mathrm{Edd}$}\label{sec:qPC_Edd}
%%%%%%%%%%%%%% The Figures %%%%%%%%%%%%%%
\begin{figure*}
    \centering
    \includegraphics[width=2.0\columnwidth]{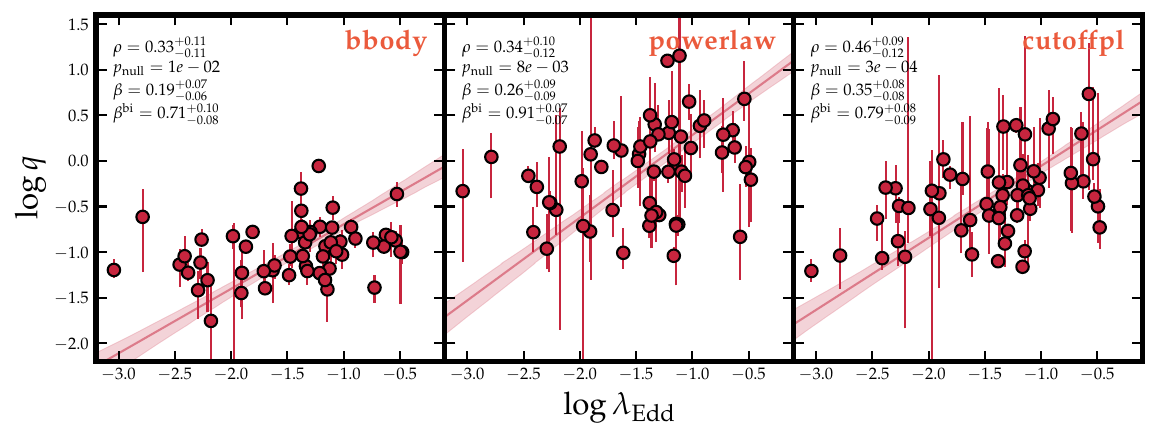}
    \caption{Relation between SE strength $q$ and Eddington ratio $\lambda_\mathrm{Edd}$ for different SE models reveals a significant correlation across all models.}
    \label{fig:qPC_Edd}
\end{figure*}
%%%%%%%%%%%%%%%%%%%%%%%%%%%%%%%%%%%%%%%%%
%%%%%%%%%%%%%% The Tables %%%%%%%%%%%%%%%
\begin{deluxetable}{lcc}
\tabletypesize{\footnotesize}
\renewcommand{\arraystretch}{1.4}
\setlength{\tabcolsep}{2.5pt}
\tablecaption{Correlation coefficients between SE strength (relative to X-rays or UV) and proxies for Eddington ratio (estimated from UV or X-rays). \label{tab:qPC_Edd}}
\tablewidth{0pt}
\tablehead{
\colhead{} &
\colhead{$\tfrac{L_\mathrm{UV}}{L_\mathrm{Edd}}=\tfrac{1}{2.75}\,\lambda_\mathrm{Edd}$} &
\colhead{$\tfrac{L_\mathrm{PC,0.5-10}}{L_\mathrm{Edd}}$}
}
\startdata
\ $\tfrac{L_\mathrm{SE,0.5-2}}{L_\mathrm{PC,0.5-2}}$ & strong (\asymunc{0.46}{0.12}{0.09}) & weak (\asymunc{0.15}{0.14}{0.13}) \\
\ $\tfrac{L_\mathrm{SE,0.5-2}}{L_\mathrm{UV}}$ & weak (\asymunc{0.03}{0.13}{0.12}) & mild (\asymunc{0.25}{0.13}{0.13}) \\
\enddata
\end{deluxetable}
%%%%%%%%%%%%%%%%%%%%%%%%%%%%%%%%%%%%%%%%%
Our data reveal a robust correlation between $q$ and $\lambda_\mathrm{Edd}$, independent of the SE model used (Fig. \ref{fig:qPC_Edd}). This well-established trend has also been demonstrated in earlier works employing both phenomenological \citep[e.g.,][]{Boissay+2016,Gliozzi&Williams2020} and physically motivated models \citep[e.g.,][]{Palit+2024,Ballantyne+2024} to characterize SE. The most direct interpretation is that SE becomes increasingly important relative to the hot corona at higher accretion rates. The availability of UV photometry allows us to probe this further. Interestingly, when characterizing SE strength by $L_\mathrm{SE,0.5-2}/L_\mathrm{UV}$, we find its correlation with $\lambda_\mathrm{Edd}$ vanishes ($\rho=\asymunc{0.03}{0.13}{0.12}$, $p_\mathrm{null}=0.53$).

Our Eddington ratio ($\lambda_\mathrm{Edd}$) equals 2.75 times the Eddington-scaled UV luminosity. For comparison, we also consider the Eddington-scaled PC luminosity, $L_\mathrm{PC,0.5-10}/L_\mathrm{Edd}$, as an alternative proxy for Eddington ratio; see Appendix \ref{sec:Edd_X}. In this case, the correlation between SE strength $q$ and Eddington-scaled PC luminosity is $\sim$\numrange[range-phrase=--]{1}{2}$\sigma$ weaker, with $\rho\sim\num{0.15}$ (Fig. \ref{fig:qPC_PCidx_Edd_x}). Meanwhile, the SE strength relative to UV ($L_\mathrm{SE,0.5-2}/L_\mathrm{UV}$) now shows a mild correlation of \asymunc{0.25}{0.13}{0.13} with Eddington-scaled PC luminosity. A summary of these correlations is provided in Table \ref{tab:qPC_Edd} for clarity.

We interpret these results as follows:
\begin{enumerate}
    \item There is an intrinsic positive correlation between $q=L_\mathrm{SE,0.5-2}/L_\mathrm{PC,0.5-2}$ and the true Eddington ratio $\lambda_{\mathrm{Edd},true}$. This is reflected already in the $q$--$\lambda_\mathrm{Edd}$ relation (Fig. \ref{fig:qPC_Edd}). The apparent weakening of the trend when comparing with Eddington-scaled PC luminosity likely arises from the ``$X\sim1/X$ effect'', i.e., the introduction of anti-correlated noise between $1/L_\mathrm{PC,0.5-2}$ (appearing in $q$) and $L_\mathrm{PC,0.5-10}$ (appearing in $L_\mathrm{PC,0.5-10}/L_\mathrm{Edd}$).

    \item The strength of SE relative to UV ($L_\mathrm{SE,0.5-2}/L_\mathrm{UV}$) is expected to increase with $\lambda_{\mathrm{Edd},true}$, as hinted by the mild correlation between $L_\mathrm{SE,0.5-2}/L_\mathrm{UV}$ and $L_\mathrm{PC,0.5-10}/L_\mathrm{Edd}$. This trend may reflect changes in the warm corona SED with accretion rate \citep[e.g.,][]{Hagen+2024,Kang+2024,Chen+2025b}. However, when the UV-based Eddington ratio, $\lambda_\mathrm{Edd}=2.75\times L_\mathrm{UV}/L_\mathrm{Edd}$, is adopted, the intrinsic correlation is diluted because source-to-source scatter in $L_\mathrm{UV}$ introduces an artificial anti-correlation between $1/L_\mathrm{UV}$ (in $L_\mathrm{SE,0.5-2}/L_\mathrm{UV}$) and $L_\mathrm{UV}$ (in $\lambda_\mathrm{Edd}$).

    \item Combining these considerations, we infer that the correlation between $q$ and the true Eddington ratio---reported both here and previous studies---likely arises from two effects: (A) the well-known decline of X-ray loudness---defined as the ratio of the X-ray primary continuum to the UV ($\sim$bolometric) luminosity---with increasing Eddington ratio \citep[e.g.,][]{Vasudevan&Fabian2009,Netzer2019,Duras+2020,Ballantyne+2024}; and (B) an intrinsic change in the warm corona SED, manifested as an increase of SE relative to UV with Eddington ratio.
\end{enumerate}

\subsubsection{$q$ vs. $\Gamma_\mathrm{PC}$}
%%%%%%%%%%%%%% The Figures %%%%%%%%%%%%%%
\begin{figure*}
    \centering
    \includegraphics[width=2.0\columnwidth]{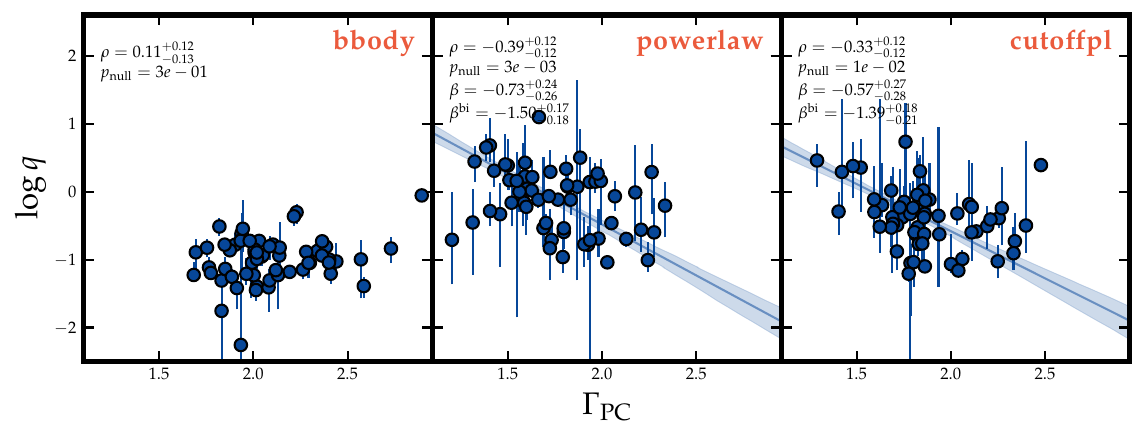}
    \caption{Relation between $q$ and PC photon index $\Gamma_\mathrm{PC}$ for different SE models. No correlation is observed when SE is fitted with a blackbody (left panel). However, a mild anti-correlation emerges for the power-law model (middle) and the cut-off power-law model (right).}
    \label{fig:qPC_PCidx}
\end{figure*}
%%%%%%%%%%%%%%%%%%%%%%%%%%%%%%%%%%%%%%%%%
The relation between SE strength $q$ and PC photon index $\Gamma_\mathrm{PC}$ depends however on the SE model, as both $q$ and $\Gamma_\mathrm{PC}$ are model sensitive (see Appendix \ref{sec:q_pcidx_bbpocpl} for further comparisons). For blackbody fits, we find no significant correlation (left panel of Fig. \ref{fig:qPC_PCidx}), consistent with earlier studies using similar models \citep[e.g.,][]{Boissay+2016,Gliozzi&Williams2020,Waddell&Gallo2020}. However, when SE is fitted with a power-law or cut-off power-law model, an anti-correlation emerges (but likely saturates at $\Gamma_\mathrm{PC}$ $>$ 2.0). This anti-correlation aligns with \citet{Palit+2024}, who employed more physically-motivated warm corona models, as well as with similar studies focusing on PG quasars \citep[e.g.,][]{Piconcelli+2005} and highly accreting SMBHs \citep[e.g.,][]{Laurenti+2024}. As demonstrated in \citetalias{Chen+2025a}, neither a blackbody nor a power-law model alone is sufficient to fully characterize the SE shape in our sample, whereas a cut-off power-law provides a more unbiased approach. Consequently, the mild anti-correlation between between $q$ and $\Gamma_\mathrm{PC}$ ($\rho=\asymunc{-0.33}{0.12}{0.12}$) observed in the right panel of Fig. \ref{fig:qPC_PCidx} is likely the most reliable. Furthermore, following the analysis in \S5.1 of \citetalias{Chen+2025a}, we confirm through simulations that this anti-correlation cannot be attributed to degeneracies between $q$ and $\Gamma_\mathrm{PC}$ introduced during X-ray spectral fitting. Notably, replacing $q$ with $L_\mathrm{SE,0.5-2}/L_\mathrm{UV}$ strengthens the anti-correlation ($\rho=\asymunc{-0.42}{0.12}{0.12}$). Since $\Gamma_\mathrm{PC}$ may also correlate with $\lambda_\mathrm{Edd}$ (see discussion in \S\ref{sec:GammaLambda}), we performed a partial correlation analysis to control for the effect of $\lambda_\mathrm{Edd}$. The resulting intrinsic anti-correlation coefficient between $q$ and $\Gamma_\mathrm{PC}$ further increases to $\rho=\asymunc{-0.56}{0.09}{0.12}$, indicating a stronger underlying relationship.

What does such anti-correlation tell us? 
Assuming the warm corona is the dominant source of SE (\S \ref{sec:3Dlum}), this anti-correlation hints a physical coupling between the hot and warm coronae. Notably, \citet{Palit+2024} recently found that the heating of the warm corona weakens in AGNs with higher $\Gamma_\mathrm{PC}$. Producing a steeper hard X-ray spectrum, i.e., higher $\Gamma_\mathrm{PC}$, requires the hot corona to exhibit either lower temperature or reduced opacity. Utilizing \nustar~data, \citet{Kang&Wang2022} showed that AGNs with higher $\Gamma_\mathrm{PC}$ tend to have coronae with higher temperatures, which inherently results in lower hot corona opacity.
The observed anti-correlation thus implies a competitive interaction between the warm and hot coronae: when the hot corona reaches higher temperatures and lower opacity, the warm corona becomes weaker. This competitive dynamic is also supported by \citetalias{Chen+2025a}, who reported an anti-correlation between $\Gamma_\mathrm{PC}$ and SE cut-off energy, indicating that a hotter, lower-opacity corona corresponds to a cooler warm corona. It is possible that a very high-temperature hot corona may partially disrupt the warm corona, reducing its capacity to receive heating from the disk. Investigating how the emissions from the warm and hot coronae vary within individual sources would be both intriguing and instrumental in clarifying the interplay between these two components.

\subsection{Implications for $\Gamma_\mathrm{PC}$ vs. $\lambda_\mathrm{Edd}$ relation}\label{sec:GammaLambda}
%%%%%%%%%%%%%% The Figures %%%%%%%%%%%%%%
\begin{figure*}
    \centering
    \includegraphics[width=2.0\columnwidth]{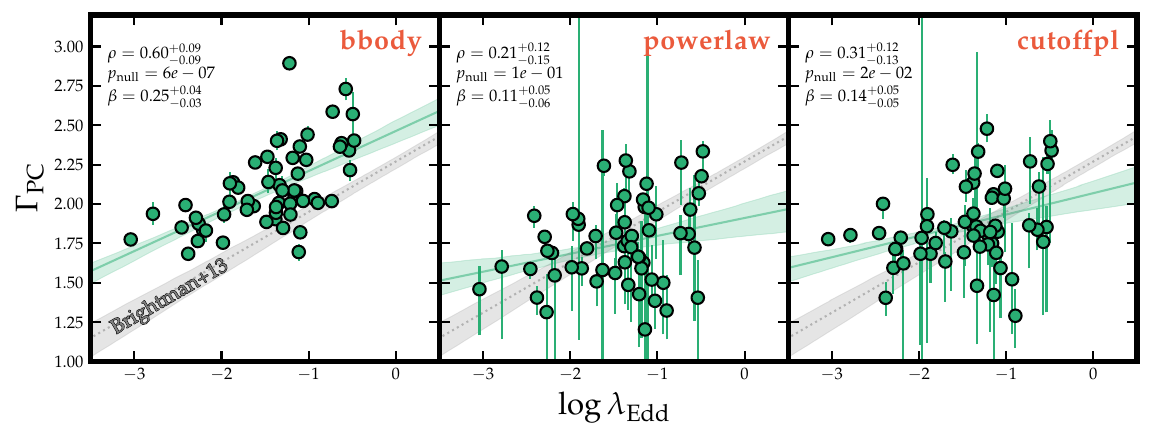}
    \caption{Relation between $\Gamma_\mathrm{PC}$ and $\lambda_\mathrm{Edd}$. Using an improper SE model, such as a blackbody, can falsely produce a strong positive correlation. When SE is instead modeled with a power-law, the correlation vanishes. The most unbiased phenomenological model---the cut-off power-law---yields only a marginal correlation ($p_\mathrm{null}=\num{2e-2}$). Furthermore, when focusing separately on sources with $\log\lambda_\mathrm{Edd}<-2$ or $>-2$, the correlation becomes insignificant in both cases. To better compare with historical literature, we present linear regressions assuming $\log\lambda_\mathrm{Edd}$ as the independent variable.}
    \label{fig:PCidx_Edd}
\end{figure*}
%%%%%%%%%%%%%%%%%%%%%%%%%%%%%%%%%%%%%%%%%
Finally, we examine the relationship between $\Gamma_\mathrm{PC}$ and $\lambda_\mathrm{Edd}$, which has been used to probe accretion disk physics \citep[e.g.,][]{Kubota&Done2018} and to measure BH mass \citep[e.g.,][]{Shemmer+2008}. Although widely cited \citep[e.g.,][]{Risaliti+2009,Brightman+2013a}, its significance has long been questioned on various grounds \citep[e.g.,][]{Trakhtenbrot+2017a,Kang&Wang2022,Waddell+2023}. Here we will further show that this correlation warrants a critical reassessment in light of the role of SE, adding an additional layer of uncertainty beyond those previously raised. For consistency with earlier work, Fig. \ref{fig:PCidx_Edd} presents the linear fit with $\log\lambda_\mathrm{Edd}$ as the independent variable (X-axis).

Historically, the hard X-ray slope $\Gamma_\mathrm{PC}$ has been measured with simple Galactic-absorbed power-law fits in the range \qtyrange[range-phrase=--,range-units=single]{2}{10}{keV}, yielding a significant correlation with $\lambda_\mathrm{Edd}$ ($p_\mathrm{null}\lesssim\num{1e-7}$) \citep[e.g.,][]{Shemmer+2008,Risaliti+2009,Brightman+2013a}. An example from \citet{Brightman+2013a} is shown as gray dashed line in Fig. \ref{fig:PCidx_Edd}. Our sample confirms this trend, showing a strong correlation between the \qtyrange[range-phrase=--,range-units=single]{2}{10}{keV} photon index (fitted with \texttt{pexrav}) and $\lambda_\mathrm{Edd}$ ($\rho=\asymunc{0.50}{0.11}{0.09}$, $p_\mathrm{null}=\num{4e-5}$). However, as noted in \citetalias{Chen+2025a}, SE contributes weakly but non-negligibly to this band ($\sim5\%$ on average for model 3), meaning that accurately deriving the intrinsic photon index of the hot corona requires incorporating the soft band and properly accounting for SE. Neglecting this component would lead to an overestimation of PC photon index by $\sim0.07$ on average for model 3.

Extending the fit to \qtyrange[range-phrase=--,range-units=single]{0.5}{10}{keV} with a blackbody SE model (model 1), we still find a strong positive correlation ($\rho=\asymunc{0.60}{0.09}{0.09}$, $p_\mathrm{null}=\num{6e-7}$, Fig. \ref{fig:PCidx_Edd}, left panel), consistent with earlier blackbody-based studies \citep[e.g.,][]{Gliozzi&Williams2020, Waddell&Gallo2020}. However, the blackbody model tends to overestimate $\Gamma_\mathrm{PC}$ when SE is strong (\citetalias{Chen+2025a}, and Appendix \ref{sec:q_pcidx_bbpocpl}), and in fact only 29\% of type 1 AGNs show blackbody-like SE, while the majority (71\%) display extended power-law-like SE. When SE is modeled with a power-law (model 2) instead, the correlation between $\Gamma_\mathrm{PC}$ and $\lambda_\mathrm{Edd}$ vanishes (middle panel of Fig. \ref{fig:PCidx_Edd}, $p_\mathrm{null}=0.1$), consistent with CAIXA sample analysis \citep{Bianchi+2009, Bianchi+2009a} where SE was similarly modeled with a power-law.

A cut-off power-law (model 3), which offers a more uniform phenomenological model capable of describing both blackbody-like and power-law-like SE, provides the least biased results. Using this model, we observe a marginal correlation ($\rho=\asymunc{0.31}{0.13}{0.12}$, $p_\mathrm{null}=0.02$; Fig. \ref{fig:PCidx_Edd}, right panel). Using the Eddington-scaled PC luminosity instead of UV-based $\lambda_\mathrm{Edd}$ gives consistent results within $1\sigma$ (Fig. \ref{fig:qPC_PCidx_Edd_x}). This correlation is however much weaker than when a simple power-law is fit to \qtyrange[range-phrase=--,range-units=single]{2}{10}{keV}, or when the soft band is incorporated with a blackbody model. Intriguingly, the data suggest a possible dividing point at $\log\lambda_\mathrm{Edd}\approx-2$: below this threshold, $\Gamma_\mathrm{PC}$ clusters around $\sim$ \num{1.75}, while above this point, $\Gamma_\mathrm{PC}$ exhibits greater scatter, with no significant correlation ($p_\mathrm{null}=0.15$). This transition aligns with the accretion state change from radiatively inefficient to radiatively efficient modes \citep[e.g.,][]{Done+2012, Yuan&Narayan2014, Ricci+2017, Hagen+2024}. While the connection between this transition and $\Gamma_\mathrm{PC}$ remains unclear, it is evident that the $\Gamma_\mathrm{PC}$--$\lambda_\mathrm{Edd}$ relationship is not a simple one-to-one continuous tight relation.

Recent studies have highlighted that the $\Gamma_\mathrm{PC}$--$\lambda_\mathrm{Edd}$ correlation breaks down when fitting spectra across higher energy bands \citep[e.g.,][]{Trakhtenbrot+2017a,Kang&Wang2022}, or when highly-accreting AGNs are considered \citep[e.g.,][]{Laurenti+2024,DegliAgosti+2025}. Our findings demonstrate that incorporating the soft band (\qtyrange[range-phrase=--,range-units=single]{0.5}{2}{keV}) and properly modeling SE also reduces the correlation's significance. These results emphasize the need for comprehensive studies incorporating both the harder X-ray band and accurate SE modeling to improve our understanding of the $\Gamma_\mathrm{PC}$--$\lambda_\mathrm{Edd}$ relation, which should be a focus of future research.

\section{Conclusions} \label{Conclusions}
In this second paper of the series, we conduct a broadband correlation analysis of 59 unobscured type 1 AGNs from \citetalias{Chen+2025a}. Our key findings are as follows:

\begin{enumerate}
    \item Our ensemble UV-to-X-ray SEDs (Fig. \ref{fig:UV_norm_spec}) reveal a strong correlation between $\alpha_\mathrm{oX}$ and the soft X-ray spectral slope $\alpha_\mathrm{0.5-2}$. Although the linear fit slope between $\alpha_\mathrm{oX}$ and $\alpha_\mathrm{0.5-2}$ deviates from a 1:1 relation, the two quantities remain directly comparable for most sources. In contrast, the hard X-ray spectral slope $\alpha_\mathrm{2-10}$ shows a weaker correlation with $\alpha_\mathrm{oX}$ and is systematically harder.

    \item We observe tight correlations between SE and UV luminosities ($\rho=\asymunc{0.92}{0.03}{0.02}$), SE and PC luminosities ($\rho=\asymunc{0.87}{0.04}{0.03}$), and PC and UV luminosities ($\rho=\asymunc{0.89}{0.04}{0.03}$) (Fig. \ref{fig:lum}), indicating couplings among all three components. Interestingly, the SE--UV relation is nearly linear, with $\beta^\mathrm{bi}=\asymunc{1.00}{0.04}{0.04}$.
    
    \item Using multiple statistical techniques, including luminosity ratio analysis (Fig. \ref{fig:lumratio}), partial correlation analysis (Fig. \ref{fig:parcorr}), and multilinear regression (Fig. \ref{fig:multireg}), we isolate the intrinsic dependence of SE on UV and PC. SE remains significantly correlated with UV even after controlling for PC, while the intrinsic SE--PC correlation (after controlling for UV) is weaker but still statistically significant. These suggest that, in addition to ionized reflection---a natural outcome of hot corona illuminating the disk---a warm corona component is essential or even dominant in producing the SE, thus favoring a hybrid scenario. This conclusion remains robust across different SE models (blackbody, power-law, cut-off power-law). 
    
    \item We find a significant correlation between SE strength $q$---defined as the ratio of SE to PC luminosity in the \qtyrange[range-phrase=--,range-units=single]{0.5}{2}{keV} band---and the Eddington ratio $\lambda_\mathrm{Edd}$ (Fig. \ref{fig:qPC_Edd}), consistent with previous studies. This trend likely arises from two concurrent effects: the well-known decline in X-ray loudness with increasing Eddington ratio, and a simultaneous rise in $L_\mathrm{SE,0.5-2}/L_\mathrm{UV}$, possibly reflecting changes in the warm corona SED.
    
    \item Additionally, we observe a mild anti-correlation between $q$ and the PC photon index $\Gamma_\mathrm{PC}$ (Fig. \ref{fig:qPC_PCidx}), likely indicating a competitive interaction between the warm and hot corona.
    
    \item Proper modeling of SE is crucial for understanding the debated correlation between $\Gamma_\mathrm{PC}$ and $\lambda_\mathrm{Edd}$ (Fig. \ref{fig:PCidx_Edd}). Different SE models yield different values of $\Gamma_\mathrm{PC}$, leading to divergent interpretations---ranging from a strong positive correlation (modeling SE with blackbody) to no correlation (modeling SE with power-law). Since the cut-off power-law model provides a more reliable treatment than either alternative, we conclude that only a marginal correlation exists. A comprehensive study incorporating both hard X-ray coverage and careful SE modeling is essential for comprehensive understanding.
\end{enumerate}

\section*{Acknowledgements}
This work was supported by the National Natural Science Foundation of China (grant nos. 12033006, 12533006, 124B1007, 12192221, 12373016, 123B2042) and the Cyrus Chun Ying Tang Foundations.

We are grateful to the referee for the thorough review and invaluable feedback. We thank Johannes Buchner from Max-Planck Institute for Extraterrestrial Physics (MPE), Shi-Fu Zhu and Jia-Lai Wang from University of Science and Technology of China (USTC) for discussions on statistical methodologies, Wen-Ke Ren from USTC for suggestions on UV data analysis, Chris Done from Durham University, Biswaraj Palit from Nicolaus Copernicus Astronomical Center, Fu-Guo Xie from Shanghai Astronomical Observatory (SHAO),  Xiaogu Zhong from Qujing Normal University, Xiaofeng Li from Changzhou Institute of Technology, and Xue-Bing Wu from Peking University for helpful scientific discussions.

The research utilized observations obtained with \xmm, an ESA science mission supported by contributions from ESA Member States and NASA. Additionally, we acknowledge the NASA/IPAC Extragalactic Database (NED), operated by the California Institute of Technology under contract with NASA \citep{NASA/IPACExtragalacticDatabaseNED2019}, for providing valuable data resources.

For bibliography, this research has made use of NASA's Astrophysics Data System, along with the adstex bibliography tool (\url{https://github.com/yymao/adstex}).

\software{
HEAsoft (v6.28; HEASARC 2014),
\xspec~\citep{Arnaud1996},
\asurv~\citep{Feigelson&Nelson1985,Isobe+1986},
\topcat~\citep{Taylor2005},
GNU Parallel Tool \citep{Tange2011}
}

% Appendix
\appendix
\section{Impact of SE model choice on spectral parameters}\label{sec:q_pcidx_bbpocpl}
\begin{figure}
    \centering
    \includegraphics[width=1.0\columnwidth]{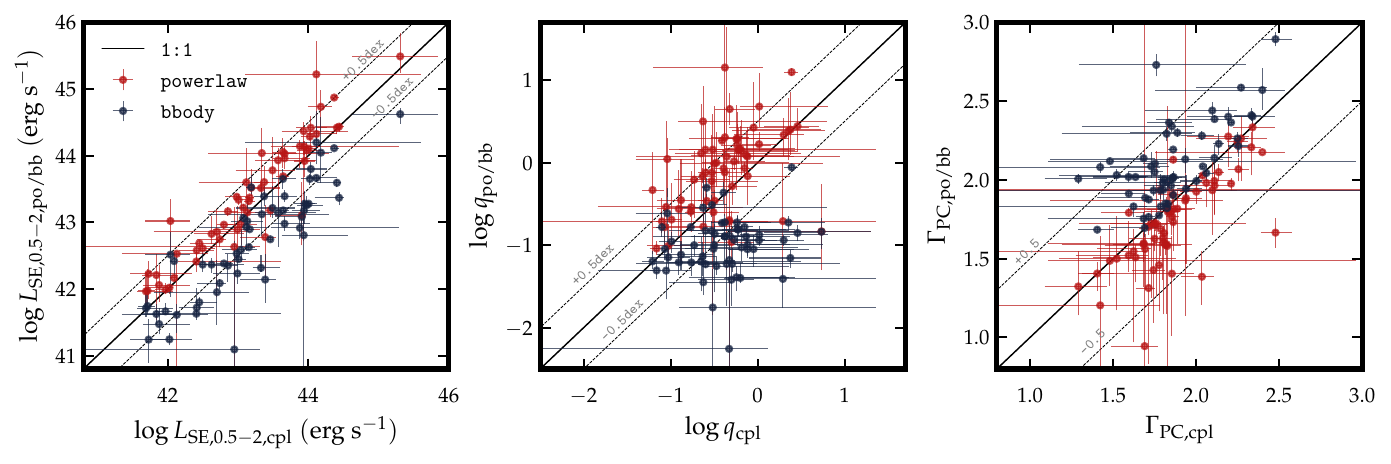}
    \caption{Comparison of SE luminosity (left), SE strength (middle), and PC photon index (right) derived from different SE models. X-axes show results from \texttt{zcutoffpl}, while Y-axes show results from \texttt{zbbody} (blue) and \texttt{zpowerlw} (red). The black line marks the 1:1 relation; gray dashed lines show $\pm0.5$ dex (or $\pm0.5$ for photon index). Compared to \texttt{zcutoffpl}, \texttt{zbbody} underestimates SE and gives a softer PC; \texttt{zpowerlw} overestimates SE and gives a harder PC.}
    \label{fig:q_pcidx_bbpocpl}
\end{figure}

We compare in Fig. \ref{fig:q_pcidx_bbpocpl} the differences in SE luminosity (left), SE strength (middle), and PC photon index (right) when using different phenomenological SE models described in \S\ref{sec:SampDR}. Physical quantities derived using \zcutoffpl~(model 3) are shown on the X-axes, while those obtained from \zbbody~(model 1) and \zpowerlw~(model 2) are shown on the Y-axes, represented by blue and red points, respectively. In each panel, the black solid line indicates the 1:1 relation, and the grey dashed lines denote $\pm0.5\ \mathrm{dex}$ (for the left and middle panels) or $\pm0.5$ (for the right panel).

We find that, compared to \zcutoffpl, the \zbbody~model tends to underestimate SE luminosity by $\sim\qty{0.5}{dex}$. This is likely because the blackbody fails to account for the extended SE components beyond $\sim$\qtyrange[range-phrase=--,range-units=single]{1}{2}{keV} in some sources, misattributing part of SE to PC. This further leads to an artificially softened PC as shown in right panel, and consequently a more significantly underestimated SE strength (deviation $\gtrsim\qty{0.5}{dex}$) in the middle panel.

On the other hand, the \zpowerlw~model tends to overestimate SE luminosity and SE strength by up to $\sim\qty{0.5}{dex}$. Compared to a blackbody, the power-law model better captures the extended shape of SE. However, due to its rigid functional form (i.e., without a cut-off), it can sometimes lead to a harder inferred PC (right panel) and an exaggerated SE component.

\section{Relations with Eddington-scaled PC luminosity}\label{sec:Edd_X}
%%%%%%%%%%%%%% The Figures %%%%%%%%%%%%%%
\begin{figure*}
    \centering
    \includegraphics[width=0.9\linewidth]{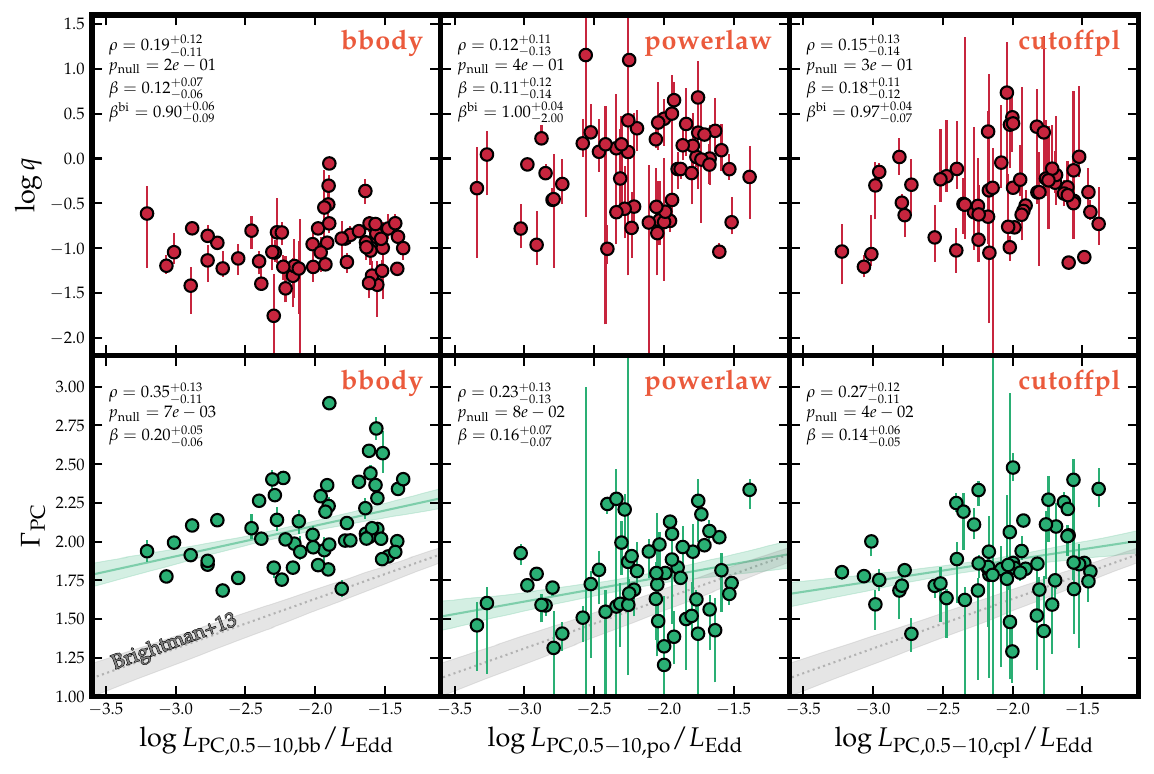}
    \caption{Same as Figs. \ref{fig:qPC_Edd} and \ref{fig:PCidx_Edd}, but with $\lambda_\mathrm{Edd}$ (Eddington-scaled UV luminosity multiplied by \num{2.75}) replaced by the Eddington-scaled PC luminosity $L_\mathrm{PC,0.5-10}/L_\mathrm{Edd}$. The $q$ correlation (top panels) weakens by $\sim$\numrange[range-phrase=--]{1}{2}$\sigma$ (vs. Fig. \ref{fig:qPC_Edd}), while the $\Gamma_\mathrm{PC}$ correlation (bottom panels) remains consistent within $1\sigma$ (vs. Fig. \ref{fig:PCidx_Edd}).}
    \label{fig:qPC_PCidx_Edd_x}
\end{figure*}
%%%%%%%%%%%%%%%%%%%%%%%%%%%%%%%%%%%%%%%%%
The $\lambda_\mathrm{Edd}$ appearing in Figs. \ref{fig:qPC_Edd} and \ref{fig:PCidx_Edd} is proportional to the Eddington-scaled UV luminosity, with a constant proportionality factor of \num{2.75}, i.e., $\lambda_\mathrm{Edd}=2.75\times L_\mathrm{UV}/L_\mathrm{Edd}$. In addition, following \citealt{Nandi+2023}, we also examine correlations between the spectral parameters (SE strength and PC photon index) with Eddington-scaled PC luminosity ($L_\mathrm{PC,0.5-10}/L_\mathrm{Edd}$).

The top panels of Fig. \ref{fig:qPC_PCidx_Edd_x} show the correlation between $q$ and Eddington-scaled PC luminosity. Compared with $q$--$\lambda_\mathrm{Edd}=2.75\times L_\mathrm{UV}/L_\mathrm{Edd}$ relation, this correlation is weaker by $\sim$\numrange[range-phrase=--]{1}{2}$\sigma$, likely due to the ``$X\sim 1/X$ effect'' discussed in \S\ref{sec:qPC_Edd}. On the other hand, the correlation between $\Gamma_\mathrm{PC}$ and $L_\mathrm{PC,0.5-10}/L_\mathrm{Edd}$ in the bottom panels remains broadly consistent with that in Fig. \ref{fig:PCidx_Edd}, differing by $\lesssim 1\sigma$.

%% For this sample we use BibTeX plus aasjournals.bst to generate the
%% the bibliography. The sample631.bib file was populated from ADS. To
%% get the citations to show in the compiled file do the following:
%%
%% pdflatex sample631.tex
%% bibtext sample631
%% pdflatex sample631.tex
%% pdflatex sample631.tex

\bibliography{XMMSE_lum}{}

\begin{thebibliography}{}
\expandafter\ifx\csname natexlab\endcsname\relax\def\natexlab#1{#1}\fi
\providecommand{\url}[1]{\href{#1}{#1}}
\providecommand{\dodoi}[1]{doi:~\href{http://doi.org/#1}{\nolinkurl{#1}}}
\providecommand{\doeprint}[1]{\href{http://ascl.net/#1}{\nolinkurl{http://ascl.net/#1}}}
\providecommand{\doarXiv}[1]{\href{https://arxiv.org/abs/#1}{\nolinkurl{https://arxiv.org/abs/#1}}}

\bibitem[{Arnaud(1996)}]{Arnaud1996}
Arnaud, K.~A. 1996, in Astronomical Society of the Pacific Conference Series,
  Vol. 101, Astronomical Data Analysis Software and Systems V, ed. G.~H. Jacoby
  \& J.~Barnes, 17

\bibitem[{{Arnaud} {et~al.}(1985){Arnaud}, {Branduardi-Raymont}, {Culhane},
  {Fabian}, {Hazard}, {McGlynn}, {Shafer}, {Tennant}, \& {Ward}}]{Arnaud+1985}
{Arnaud}, K.~A., {Branduardi-Raymont}, G., {Culhane}, J.~L., {et~al.} 1985,
  \mnras, 217, 105, \dodoi{10.1093/mnras/217.1.105}

\bibitem[{{Ballantyne}(2020)}]{Ballantyne2020}
{Ballantyne}, D.~R. 2020, \mnras, 491, 3553, \dodoi{10.1093/mnras/stz3294}

\bibitem[{{Ballantyne} {et~al.}(2001){Ballantyne}, {Ross}, \&
  {Fabian}}]{Ballantyne+2001}
{Ballantyne}, D.~R., {Ross}, R.~R., \& {Fabian}, A.~C. 2001, \mnras, 327, 10,
  \dodoi{10.1046/j.1365-8711.2001.04432.x}

\bibitem[{{Ballantyne} \& {Xiang}(2020)}]{Ballantyne&Xiang2020}
{Ballantyne}, D.~R., \& {Xiang}, X. 2020, \mnras, 496, 4255,
  \dodoi{10.1093/mnras/staa1866}

\bibitem[{{Ballantyne} {et~al.}(2024){Ballantyne}, {Sudhakar}, {Fairfax},
  {Bianchi}, {Czerny}, {De Rosa}, {De Marco}, {Middei}, {Palit}, {Petrucci},
  {R{\'o}{\.z}a{\'n}ska}, \& {Ursini}}]{Ballantyne+2024}
{Ballantyne}, D.~R., {Sudhakar}, V., {Fairfax}, D., {et~al.} 2024, \mnras, 530,
  1603, \dodoi{10.1093/mnras/stae944}

\bibitem[{{Bambi} {et~al.}(2021){Bambi}, {Brenneman}, {Dauser}, {Garc{\'\i}a},
  {Grinberg}, {Ingram}, {Jiang}, {Liu}, {Lohfink}, {Marinucci}, {Mastroserio},
  {Middei}, {Nampalliwar}, {Nied{\'z}wiecki}, {Steiner}, {Tripathi}, \&
  {Zdziarski}}]{Bambi+2021}
{Bambi}, C., {Brenneman}, L.~W., {Dauser}, T., {et~al.} 2021, \ssr, 217, 65,
  \dodoi{10.1007/s11214-021-00841-8}

\bibitem[{{Bianchi} {et~al.}(2009{\natexlab{a}}){Bianchi}, {Bonilla},
  {Guainazzi}, {Matt}, \& {Ponti}}]{Bianchi+2009}
{Bianchi}, S., {Bonilla}, N.~F., {Guainazzi}, M., {Matt}, G., \& {Ponti}, G.
  2009{\natexlab{a}}, \aap, 501, 915, \dodoi{10.1051/0004-6361/200911905}

\bibitem[{{Bianchi} {et~al.}(2009{\natexlab{b}}){Bianchi}, {Guainazzi}, {Matt},
  {Fonseca Bonilla}, \& {Ponti}}]{Bianchi+2009a}
{Bianchi}, S., {Guainazzi}, M., {Matt}, G., {Fonseca Bonilla}, N., \& {Ponti},
  G. 2009{\natexlab{b}}, \aap, 495, 421, \dodoi{10.1051/0004-6361:200810620}

\bibitem[{{Boissay} {et~al.}(2016){Boissay}, {Ricci}, \&
  {Paltani}}]{Boissay+2016}
{Boissay}, R., {Ricci}, C., \& {Paltani}, S. 2016, \aap, 588, A70,
  \dodoi{10.1051/0004-6361/201526982}

\bibitem[{{Boissay} {et~al.}(2014){Boissay}, {Paltani}, {Ponti}, {Bianchi},
  {Cappi}, {Kaastra}, {Petrucci}, {Arav}, {Branduardi-Raymont}, {Costantini},
  {Ebrero}, {Kriss}, {Mehdipour}, {Pinto}, \& {Steenbrugge}}]{Boissay+2014}
{Boissay}, R., {Paltani}, S., {Ponti}, G., {et~al.} 2014, \aap, 567, A44,
  \dodoi{10.1051/0004-6361/201423494}

\bibitem[{{Brightman} {et~al.}(2013){Brightman}, {Silverman}, {Mainieri},
  {Ueda}, {Schramm}, {Matsuoka}, {Nagao}, {Steinhardt}, {Kartaltepe},
  {Sanders}, {Treister}, {Shemmer}, {Brandt}, {Brusa}, {Comastri}, {Ho},
  {Lanzuisi}, {Lusso}, {Nandra}, {Salvato}, {Zamorani}, {Akiyama}, {Alexander},
  {Bongiorno}, {Capak}, {Civano}, {Del Moro}, {Doi}, {Elvis}, {Hasinger},
  {Laird}, {Masters}, {Mignoli}, {Ohta}, {Schawinski}, \&
  {Taniguchi}}]{Brightman+2013a}
{Brightman}, M., {Silverman}, J.~D., {Mainieri}, V., {et~al.} 2013, \mnras,
  433, 2485, \dodoi{10.1093/mnras/stt920}

\bibitem[{{Cai}(2024)}]{Cai2024}
{Cai}, Z. 2024, Universe, 10, 431, \dodoi{10.3390/universe10110431}

\bibitem[{{Cai} \& {Wang}(2023)}]{Cai&Wang2023}
{Cai}, Z.-Y., \& {Wang}, J.-X. 2023, Nature Astronomy, 7, 1506,
  \dodoi{10.1038/s41550-023-02088-5}

\bibitem[{{Chalise} {et~al.}(2022){Chalise}, {Lohfink}, {Chauhan}, {Russell},
  {Buisson}, \& {Mallick}}]{Chalise+2022}
{Chalise}, S., {Lohfink}, A.~M., {Chauhan}, J., {et~al.} 2022, \mnras, 517,
  4788, \dodoi{10.1093/mnras/stac2953}

\bibitem[{{Chen} {et~al.}(2025{\natexlab{a}}){Chen}, {Wang}, {Kang}, {Kang},
  {Sou}, {Liu}, {Cai}, \& {Su}}]{Chen+2025a}
{Chen}, S.-J., {Wang}, J.-X., {Kang}, J.-L., {et~al.} 2025{\natexlab{a}}, \apj,
  980, 23, \dodoi{10.3847/1538-4357/ada035}

\bibitem[{{Chen} {et~al.}(2025{\natexlab{b}}){Chen}, {Buchner}, {Liu}, {Hagen},
  {Waddell}, {Nandra}, {Salvato}, {Igo}, {Aydar}, {Merloni}, {Ni}, {Kang},
  {Cai}, {Wang}, {Li}, {Ramos-Ceja}, {Sanders}, {Georgakakis}, \&
  {Zhang}}]{Chen+2025b}
{Chen}, S.-J., {Buchner}, J., {Liu}, T., {et~al.} 2025{\natexlab{b}}, \aap,
  701, A144, \dodoi{10.1051/0004-6361/202554737}

\bibitem[{{Chiang} {et~al.}(2015){Chiang}, {Walton}, {Fabian}, {Wilkins}, \&
  {Gallo}}]{Chiang+2015}
{Chiang}, C.-Y., {Walton}, D.~J., {Fabian}, A.~C., {Wilkins}, D.~R., \&
  {Gallo}, L.~C. 2015, \mnras, 446, 759, \dodoi{10.1093/mnras/stu2087}

\bibitem[{{Cruise} {et~al.}(2025){Cruise}, {Guainazzi}, {Aird}, {Carrera},
  {Costantini}, {Corrales}, {Dauser}, {Eckert}, {Gastaldello}, {Matsumoto},
  {Osten}, {Petrucci}, {Porquet}, {Pratt}, {Rea}, {Reiprich}, {Simionescu},
  {Spiga}, \& {Troja}}]{Cruise+2024}
{Cruise}, M., {Guainazzi}, M., {Aird}, J., {et~al.} 2025, Nature Astronomy, 9,
  36, \dodoi{10.1038/s41550-024-02416-3}

\bibitem[{{Crummy} {et~al.}(2006){Crummy}, {Fabian}, {Gallo}, \&
  {Ross}}]{Crummy+2005}
{Crummy}, J., {Fabian}, A.~C., {Gallo}, L., \& {Ross}, R.~R. 2006, \mnras, 365,
  1067, \dodoi{10.1111/j.1365-2966.2005.09844.x}

\bibitem[{{Degli Agosti} {et~al.}(2025){Degli Agosti}, {Vignali}, {Piconcelli},
  {Zappacosta}, {Bertola}, {Middei}, {Saccheo}, {Vietri}, {Vito}, {Bongiorno},
  {Bischetti}, {Bruni}, {Carniani}, {Cresci}, {Feruglio}, {Salvestrini},
  {Travascio}, {Gaspari}, {Glikman}, {Kammoun}, {Lanzuisi}, {Laurenti},
  {Miniutti}, {Pinto}, {Testa}, {Tombesi}, {Tortosa}, \&
  {Fiore}}]{DegliAgosti+2025}
{Degli Agosti}, C., {Vignali}, C., {Piconcelli}, E., {et~al.} 2025, arXiv
  e-prints, arXiv:2509.08055.
\newblock \doarXiv{2509.08055}

\bibitem[{{Dewangan} {et~al.}(2007){Dewangan}, {Griffiths}, {Dasgupta}, \&
  {Rao}}]{Dewangan+2007}
{Dewangan}, G.~C., {Griffiths}, R.~E., {Dasgupta}, S., \& {Rao}, A.~R. 2007,
  \apj, 671, 1284, \dodoi{10.1086/523683}

\bibitem[{{Ding} {et~al.}(2024){Ding}, {Garc{\i}a}, {Kallman}, {Mendoza},
  {Bautista}, {Harrison}, {Tomsick}, \& {Dong}}]{Ding+2024}
{Ding}, Y., {Garc{\i}a}, J.~A., {Kallman}, T.~R., {et~al.} 2024, \apj, 974,
  280, \dodoi{10.3847/1538-4357/ad76a1}

\bibitem[{{Done} {et~al.}(2012){Done}, {Davis}, {Jin}, {Blaes}, \&
  {Ward}}]{Done+2012}
{Done}, C., {Davis}, S.~W., {Jin}, C., {Blaes}, O., \& {Ward}, M. 2012, \mnras,
  420, 1848, \dodoi{10.1111/j.1365-2966.2011.19779.x}

\bibitem[{{Dov{\v{c}}iak} {et~al.}(2011){Dov{\v{c}}iak}, {Muleri}, {Goosmann},
  {Karas}, \& {Matt}}]{Dovciak+2011}
{Dov{\v{c}}iak}, M., {Muleri}, F., {Goosmann}, R.~W., {Karas}, V., \& {Matt},
  G. 2011, \apj, 731, 75, \dodoi{10.1088/0004-637X/731/1/75}

\bibitem[{{Duras} {et~al.}(2020){Duras}, {Bongiorno}, {Ricci}, {Piconcelli},
  {Shankar}, {Lusso}, {Bianchi}, {Fiore}, {Maiolino}, {Marconi}, {Onori},
  {Sani}, {Schneider}, {Vignali}, \& {La Franca}}]{Duras+2020}
{Duras}, F., {Bongiorno}, A., {Ricci}, F., {et~al.} 2020, \aap, 636, A73,
  \dodoi{10.1051/0004-6361/201936817}

\bibitem[{{Emmanoulopoulos} {et~al.}(2011){Emmanoulopoulos}, {McHardy}, \&
  {Papadakis}}]{Emmanoulopoulos+2011}
{Emmanoulopoulos}, D., {McHardy}, I.~M., \& {Papadakis}, I.~E. 2011, \mnras,
  416, L94, \dodoi{10.1111/j.1745-3933.2011.01106.x}

\bibitem[{{Fabian} {et~al.}(2012){Fabian}, {Zoghbi}, {Wilkins}, {Dwelly},
  {Uttley}, {Schartel}, {Miniutti}, {Gallo}, {Grupe}, {Komossa}, \&
  {Santos-Lle{\'o}}}]{Fabian+2012}
{Fabian}, A.~C., {Zoghbi}, A., {Wilkins}, D., {et~al.} 2012, \mnras, 419, 116,
  \dodoi{10.1111/j.1365-2966.2011.19676.x}

\bibitem[{{Feigelson} \& {Nelson}(1985)}]{Feigelson&Nelson1985}
{Feigelson}, E.~D., \& {Nelson}, P.~I. 1985, \apj, 293, 192,
  \dodoi{10.1086/163225}

\bibitem[{{Garc{\'\i}a} \& {Kallman}(2010)}]{Garcia&Kallman2010}
{Garc{\'\i}a}, J., \& {Kallman}, T.~R. 2010, \apj, 718, 695,
  \dodoi{10.1088/0004-637X/718/2/695}

\bibitem[{{Garc{\'\i}a} {et~al.}(2014){Garc{\'\i}a}, {Dauser}, {Lohfink},
  {Kallman}, {Steiner}, {McClintock}, {Brenneman}, {Wilms}, {Eikmann},
  {Reynolds}, \& {Tombesi}}]{Garcia+2014}
{Garc{\'\i}a}, J., {Dauser}, T., {Lohfink}, A., {et~al.} 2014, \apj, 782, 76,
  \dodoi{10.1088/0004-637X/782/2/76}

\bibitem[{{Garc{\'\i}a} {et~al.}(2016){Garc{\'\i}a}, {Fabian}, {Kallman},
  {Dauser}, {Parker}, {McClintock}, {Steiner}, \& {Wilms}}]{Garcia+2016}
{Garc{\'\i}a}, J.~A., {Fabian}, A.~C., {Kallman}, T.~R., {et~al.} 2016, \mnras,
  462, 751, \dodoi{10.1093/mnras/stw1696}

\bibitem[{{Garc{\'\i}a} {et~al.}(2019){Garc{\'\i}a}, {Kara}, {Walton},
  {Beuchert}, {Dauser}, {Gatuzz}, {Balokovic}, {Steiner}, {Tombesi}, {Connors},
  {Kallman}, {Harrison}, {Fabian}, {Wilms}, {Stern}, {Lanz}, {Ricci}, \&
  {Ballantyne}}]{Garcia+2019}
{Garc{\'\i}a}, J.~A., {Kara}, E., {Walton}, D., {et~al.} 2019, \apj, 871, 88,
  \dodoi{10.3847/1538-4357/aaf739}

\bibitem[{{George} {et~al.}(2000){George}, {Turner}, {Yaqoob}, {Netzer},
  {Laor}, {Mushotzky}, {Nandra}, \& {Takahashi}}]{George+2000}
{George}, I.~M., {Turner}, T.~J., {Yaqoob}, T., {et~al.} 2000, \apj, 531, 52,
  \dodoi{10.1086/308461}

\bibitem[{{Gliozzi} \& {Williams}(2020)}]{Gliozzi&Williams2020}
{Gliozzi}, M., \& {Williams}, J.~K. 2020, \mnras, 491, 532,
  \dodoi{10.1093/mnras/stz3005}

\bibitem[{{Gronkiewicz} \&
  {R{\'o}{\.z}a{\'n}ska}(2020)}]{Gronkiewicz&Rozanska2020}
{Gronkiewicz}, D., \& {R{\'o}{\.z}a{\'n}ska}, A. 2020, \aap, 633, A35,
  \dodoi{10.1051/0004-6361/201935033}

\bibitem[{{Grupe} {et~al.}(2010){Grupe}, {Komossa}, {Leighly}, \&
  {Page}}]{Grupe+2010}
{Grupe}, D., {Komossa}, S., {Leighly}, K.~M., \& {Page}, K.~L. 2010, \apjs,
  187, 64, \dodoi{10.1088/0067-0049/187/1/64}

\bibitem[{{Hagen} {et~al.}(2024){Hagen}, {Done}, {Silverman}, {Li}, {Liu},
  {Ren}, {Buchner}, {Merloni}, {Nagao}, \& {Salvato}}]{Hagen+2024}
{Hagen}, S., {Done}, C., {Silverman}, J.~D., {et~al.} 2024, \mnras, 534, 2803,
  \dodoi{10.1093/mnras/stae2272}

\bibitem[{{Hancock} {et~al.}(2022){Hancock}, {Young}, \&
  {Chainakun}}]{Hancock+2022}
{Hancock}, S., {Young}, A.~J., \& {Chainakun}, P. 2022, \mnras, 514, 5403,
  \dodoi{10.1093/mnras/stac1653}

\bibitem[{{Isobe} {et~al.}(1990){Isobe}, {Feigelson}, {Akritas}, \&
  {Babu}}]{Isobe&Feigelson1990}
{Isobe}, T., {Feigelson}, E.~D., {Akritas}, M.~G., \& {Babu}, G.~J. 1990, \apj,
  364, 104, \dodoi{10.1086/169390}

\bibitem[{{Isobe} {et~al.}(1986){Isobe}, {Feigelson}, \& {Nelson}}]{Isobe+1986}
{Isobe}, T., {Feigelson}, E.~D., \& {Nelson}, P.~I. 1986, \apj, 306, 490,
  \dodoi{10.1086/164359}

\bibitem[{{Jiang} {et~al.}(2018){Jiang}, {Parker}, {Fabian}, {Alston},
  {Buisson}, {Cackett}, {Chiang}, {Dauser}, {Gallo}, {Garc{\'\i}a}, {Harrison},
  {Lohfink}, {De Marco}, {Kara}, {Miller}, {Miniutti}, {Pinto}, {Walton}, \&
  {Wilkins}}]{Jiang+2018}
{Jiang}, J., {Parker}, M.~L., {Fabian}, A.~C., {et~al.} 2018, \mnras, 477,
  3711, \dodoi{10.1093/mnras/sty836}

\bibitem[{{Jiang} {et~al.}(2019){Jiang}, {Fabian}, {Dauser}, {Gallo},
  {Garc{\'\i}a}, {Kara}, {Parker}, {Tomsick}, {Walton}, \&
  {Reynolds}}]{Jiang+2019}
{Jiang}, J., {Fabian}, A.~C., {Dauser}, T., {et~al.} 2019, \mnras, 489, 3436,
  \dodoi{10.1093/mnras/stz2326}

\bibitem[{{Jin} {et~al.}(2016){Jin}, {Done}, \& {Ward}}]{Jin+2016}
{Jin}, C., {Done}, C., \& {Ward}, M. 2016, \mnras, 455, 691,
  \dodoi{10.1093/mnras/stv2319}

\bibitem[{{Kang} {et~al.}(2020){Kang}, {Wang}, \& {Kang}}]{Kang2020}
{Kang}, J., {Wang}, J., \& {Kang}, W. 2020, \apj, 901, 111,
  \dodoi{10.3847/1538-4357/abadf5}

\bibitem[{{Kang} {et~al.}(2025){Kang}, {Done}, {Hagen}, {Temple}, {Silverman},
  {Li}, \& {Liu}}]{Kang+2024}
{Kang}, J.-L., {Done}, C., {Hagen}, S., {et~al.} 2025, \mnras, 538, 121,
  \dodoi{10.1093/mnras/staf145}

\bibitem[{{Kang} \& {Wang}(2022)}]{Kang&Wang2022}
{Kang}, J.-L., \& {Wang}, J.-X. 2022, \apj, 929, 141,
  \dodoi{10.3847/1538-4357/ac5d49}

\bibitem[{{Kara} {et~al.}(2013){Kara}, {Fabian}, {Cackett}, {Steiner},
  {Uttley}, {Wilkins}, \& {Zoghbi}}]{Kara+2013}
{Kara}, E., {Fabian}, A.~C., {Cackett}, E.~M., {et~al.} 2013, \mnras, 428,
  2795, \dodoi{10.1093/mnras/sts155}

\bibitem[{{Kara} \& {Garc{\'\i}a}(2025)}]{Kara&Garcia2025}
{Kara}, E., \& {Garc{\'\i}a}, J. 2025, arXiv e-prints, arXiv:2503.22791,
  \dodoi{10.48550/arXiv.2503.22791}

\bibitem[{{Kawanaka} \& {Mineshige}(2024)}]{Kawanaka&Mineshige2024}
{Kawanaka}, N., \& {Mineshige}, S. 2024, \pasj, 76, 306,
  \dodoi{10.1093/pasj/psae012}

\bibitem[{{Krawczyk} {et~al.}(2013){Krawczyk}, {Richards}, {Mehta}, {Vogeley},
  {Gallagher}, {Leighly}, {Ross}, \& {Schneider}}]{Krawczyk+2013}
{Krawczyk}, C.~M., {Richards}, G.~T., {Mehta}, S.~S., {et~al.} 2013, \apjs,
  206, 4, \dodoi{10.1088/0067-0049/206/1/4}

\bibitem[{{Kubota} \& {Done}(2018)}]{Kubota&Done2018}
{Kubota}, A., \& {Done}, C. 2018, \mnras, 480, 1247,
  \dodoi{10.1093/mnras/sty1890}

\bibitem[{{Laurenti} {et~al.}(2024){Laurenti}, {Tombesi}, {Vagnetti},
  {Piconcelli}, {Guainazzi}, {Vignali}, {Paolillo}, {Middei}, {Bongiorno}, \&
  {Zappacosta}}]{Laurenti+2024}
{Laurenti}, M., {Tombesi}, F., {Vagnetti}, F., {et~al.} 2024, \aap, 689, A337,
  \dodoi{10.1051/0004-6361/202449147}

\bibitem[{{Liu} {et~al.}(2020){Liu}, {Wang}, {Abdikamalov}, {Ayzenberg}, \&
  {Bambi}}]{Liu+2020}
{Liu}, H., {Wang}, H., {Abdikamalov}, A.~B., {Ayzenberg}, D., \& {Bambi}, C.
  2020, \apj, 896, 160, \dodoi{10.3847/1538-4357/ab917a}

\bibitem[{{Liu} \& {Qiao}(2010)}]{Liu&Qiao2010}
{Liu}, J., \& {Qiao}, E. 2010, Science China Physics, Mechanics, and Astronomy,
  53, 102, \dodoi{10.1007/s11433-010-0033-1}

\bibitem[{{Lusso} {et~al.}(2010){Lusso}, {Comastri}, {Vignali}, {Zamorani},
  {Brusa}, {Gilli}, {Iwasawa}, {Salvato}, {Civano}, {Elvis}, {Merloni},
  {Bongiorno}, {Trump}, {Koekemoer}, {Schinnerer}, {Le Floc'h}, {Cappelluti},
  {Jahnke}, {Sargent}, {Silverman}, {Mainieri}, {Fiore}, {Bolzonella}, {Le
  F{\`e}vre}, {Garilli}, {Iovino}, {Kneib}, {Lamareille}, {Lilly}, {Mignoli},
  {Scodeggio}, \& {Vergani}}]{Lusso+2010}
{Lusso}, E., {Comastri}, A., {Vignali}, C., {et~al.} 2010, \aap, 512, A34,
  \dodoi{10.1051/0004-6361/200913298}

\bibitem[{{Magdziarz} {et~al.}(1998){Magdziarz}, {Blaes}, {Zdziarski},
  {Johnson}, \& {Smith}}]{Magdziarz+1998}
{Magdziarz}, P., {Blaes}, O.~M., {Zdziarski}, A.~A., {Johnson}, W.~N., \&
  {Smith}, D.~A. 1998, \mnras, 301, 179,
  \dodoi{10.1046/j.1365-8711.1998.02015.x}

\bibitem[{{Magdziarz} \& {Zdziarski}(1995)}]{Magdziarz&Zdziarski1995}
{Magdziarz}, P., \& {Zdziarski}, A.~A. 1995, \mnras, 273, 837,
  \dodoi{10.1093/mnras/273.3.837}

\bibitem[{{Mallick} {et~al.}(2025){Mallick}, {Pinto}, {Tomsick}, {Markowitz},
  {Fabian}, {Safi-Harb}, {Steiner}, {Pacucci}, \& {Alston}}]{Mallick+2025}
{Mallick}, L., {Pinto}, C., {Tomsick}, J., {et~al.} 2025, arXiv e-prints,
  arXiv:2501.15380, \dodoi{10.48550/arXiv.2501.15380}

\bibitem[{{Mehdipour} {et~al.}(2023){Mehdipour}, {Kriss}, {Kaastra},
  {Costantini}, \& {Mao}}]{Mehdipour+2023}
{Mehdipour}, M., {Kriss}, G.~A., {Kaastra}, J.~S., {Costantini}, E., \& {Mao},
  J. 2023, \apjl, 952, L5, \dodoi{10.3847/2041-8213/ace053}

\bibitem[{Mehdipour {et~al.}(2011)Mehdipour, {Branduardi-Raymont}, Kaastra,
  Petrucci, Kriss, Ponti, Blustin, Paltani, Cappi, Detmers, \&
  Steenbrugge}]{Mehdipour+2011}
Mehdipour, M., {Branduardi-Raymont}, G., Kaastra, J.~S., {et~al.} 2011

\bibitem[{{Merloni} {et~al.}(2006){Merloni}, {Malzac}, {Fabian}, \&
  {Ross}}]{Merloni+2006}
{Merloni}, A., {Malzac}, J., {Fabian}, A.~C., \& {Ross}, R.~R. 2006, \mnras,
  370, 1699, \dodoi{10.1111/j.1365-2966.2006.10676.x}

\bibitem[{{Middei} {et~al.}(2020){Middei}, {Petrucci}, {Bianchi}, {Ursini},
  {Cappi}, {Clavel}, {De Rosa}, {Marinucci}, {Matt}, \&
  {Tortosa}}]{Middei+2020}
{Middei}, R., {Petrucci}, P.~O., {Bianchi}, S., {et~al.} 2020, \aap, 640, A99,
  \dodoi{10.1051/0004-6361/202038112}

\bibitem[{{Mineo} {et~al.}(2000){Mineo}, {Fiore}, {Laor}, {Costantini},
  {Brandt}, {Comastri}, {Della Ceca}, {Elvis}, {Maccacaro}, \&
  {Molendi}}]{Mineo+2000}
{Mineo}, T., {Fiore}, F., {Laor}, A., {et~al.} 2000, \aap, 359, 471,
  \dodoi{10.48550/arXiv.astro-ph/0005567}

\bibitem[{{Miniutti} \& {Fabian}(2004)}]{Miniutti&Fabian2004}
{Miniutti}, G., \& {Fabian}, A.~C. 2004, \mnras, 349, 1435,
  \dodoi{10.1111/j.1365-2966.2004.07611.x}

\bibitem[{{Nandi} {et~al.}(2021){Nandi}, {Chatterjee}, {Chakrabarti}, \&
  {Dutta}}]{Nandi+2021}
{Nandi}, P., {Chatterjee}, A., {Chakrabarti}, S.~K., \& {Dutta}, B.~G. 2021,
  \mnras, 506, 3111, \dodoi{10.1093/mnras/stab1699}

\bibitem[{{Nandi} {et~al.}(2023){Nandi}, {Chatterjee}, {Jana}, {Chakrabarti},
  {Naik}, {Safi-Harb}, {Chang}, \& {Heyl}}]{Nandi+2023}
{Nandi}, P., {Chatterjee}, A., {Jana}, A., {et~al.} 2023, \apjs, 269, 15,
  \dodoi{10.3847/1538-4365/acf4f9}

\bibitem[{{NASA/IPAC Extragalactic Database
  (NED)}(2019)}]{NASA/IPACExtragalacticDatabaseNED2019}
{NASA/IPAC Extragalactic Database (NED)}. 2019, IPAC, \dodoi{10.26132/NED1}

\bibitem[{{Natali} {et~al.}(1998){Natali}, {Giallongo}, {Cristiani}, \& {La
  Franca}}]{Natali+1998}
{Natali}, F., {Giallongo}, E., {Cristiani}, S., \& {La Franca}, F. 1998, \aj,
  115, 397, \dodoi{10.1086/300211}

\bibitem[{{Netzer}(2019)}]{Netzer2019}
{Netzer}, H. 2019, \mnras, 488, 5185, \dodoi{10.1093/mnras/stz2016}

\bibitem[{{Noda} \& {Done}(2018)}]{Noda&Done2018}
{Noda}, H., \& {Done}, C. 2018, \mnras, 480, 3898,
  \dodoi{10.1093/mnras/sty2032}

\bibitem[{{Nour} \& {Sriram}(2023)}]{Nour&Sriram2022}
{Nour}, D., \& {Sriram}, K. 2023, \mnras, 518, 5705,
  \dodoi{10.1093/mnras/stac3505}

\bibitem[{{Palit} {et~al.}(2024){Palit}, {R{\'o}{\.z}a{\'n}ska}, {Petrucci},
  {Gronkiewicz}, {Barnier}, {Bianchi}, {Ballantyne}, {Gianolli}, {Middei},
  {Belmont}, \& {Ursini}}]{Palit+2024}
{Palit}, B., {R{\'o}{\.z}a{\'n}ska}, A., {Petrucci}, P.~O., {et~al.} 2024,
  \aap, 690, A308, \dodoi{10.1051/0004-6361/202450111}

\bibitem[{{Partington} {et~al.}(2024){Partington}, {Cackett}, {Edelson},
  {Horne}, {Gelbord}, {Kara}, {Malacaria}, {Miller}, {Steiner}, \&
  {Sanna}}]{Partington+2024}
{Partington}, E.~R., {Cackett}, E.~M., {Edelson}, R., {et~al.} 2024, \apj, 977,
  77, \dodoi{10.3847/1538-4357/ad8dc2}

\bibitem[{Pedregosa {et~al.}(2011)Pedregosa, Varoquaux, Gramfort, Michel,
  Thirion, Grisel, Blondel, Prettenhofer, Weiss, Dubourg, Vanderplas, Passos,
  Cournapeau, Brucher, Perrot, \& Duchesnay}]{Pedregosa+2011}
Pedregosa, F., Varoquaux, G., Gramfort, A., {et~al.} 2011, Journal of Machine
  Learning Research, 12, 2825

\bibitem[{{Petrucci} {et~al.}(2018){Petrucci}, {Ursini}, {De Rosa}, {Bianchi},
  {Cappi}, {Matt}, {Dadina}, \& {Malzac}}]{Petrucci+2018}
{Petrucci}, P.~O., {Ursini}, F., {De Rosa}, A., {et~al.} 2018, \aap, 611, A59,
  \dodoi{10.1051/0004-6361/201731580}

\bibitem[{{Petrucci} {et~al.}(2013){Petrucci}, {Paltani}, {Malzac}, {Kaastra},
  {Cappi}, {Ponti}, {De Marco}, {Kriss}, {Steenbrugge}, {Bianchi},
  {Branduardi-Raymont}, {Mehdipour}, {Costantini}, {Dadina}, \&
  {Lubi{\'n}ski}}]{Petrucci+2013}
{Petrucci}, P.~O., {Paltani}, S., {Malzac}, J., {et~al.} 2013, \aap, 549, A73,
  \dodoi{10.1051/0004-6361/201219956}

\bibitem[{{Petrucci} {et~al.}(2020){Petrucci}, {Gronkiewicz}, {Rozanska},
  {Belmont}, {Bianchi}, {Czerny}, {Matt}, {Malzac}, {Middei}, {De Rosa},
  {Ursini}, \& {Cappi}}]{Petrucci+2020}
{Petrucci}, P.~O., {Gronkiewicz}, D., {Rozanska}, A., {et~al.} 2020, \aap, 634,
  A85, \dodoi{10.1051/0004-6361/201937011}

\bibitem[{{Piconcelli} {et~al.}(2005){Piconcelli}, {Jimenez-Bail{\'o}n},
  {Guainazzi}, {Schartel}, {Rodr{\'\i}guez-Pascual}, \&
  {Santos-Lle{\'o}}}]{Piconcelli+2005}
{Piconcelli}, E., {Jimenez-Bail{\'o}n}, E., {Guainazzi}, M., {et~al.} 2005,
  \aap, 432, 15, \dodoi{10.1051/0004-6361:20041621}

\bibitem[{{Porquet} {et~al.}(2025){Porquet}, {Reeves}, \&
  {Braito}}]{Porquet+2025}
{Porquet}, D., {Reeves}, J.~N., \& {Braito}, V. 2025, arXiv e-prints,
  arXiv:2506.23920, \dodoi{10.48550/arXiv.2506.23920}

\bibitem[{{Porquet} {et~al.}(2004){Porquet}, {Reeves}, {O'Brien}, \&
  {Brinkmann}}]{Porquet+2004}
{Porquet}, D., {Reeves}, J.~N., {O'Brien}, P., \& {Brinkmann}, W. 2004, \aap,
  422, 85, \dodoi{10.1051/0004-6361:20047108}

\bibitem[{{Porquet} {et~al.}(2018){Porquet}, {Reeves}, {Matt}, {Marinucci},
  {Nardini}, {Braito}, {Lobban}, {Ballantyne}, {Boggs}, {Christensen},
  {Dauser}, {Farrah}, {Garcia}, {Hailey}, {Harrison}, {Stern}, {Tortosa},
  {Ursini}, \& {Zhang}}]{Porquet+2018}
{Porquet}, D., {Reeves}, J.~N., {Matt}, G., {et~al.} 2018, \aap, 609, A42,
  \dodoi{10.1051/0004-6361/201731290}

\bibitem[{{Pravdo} {et~al.}(1981){Pravdo}, {Nugent}, {Nousek}, {Jensen},
  {Wilson}, \& {Becker}}]{Pravdo+1981}
{Pravdo}, S.~H., {Nugent}, J.~J., {Nousek}, J.~A., {et~al.} 1981, \apj, 251,
  501, \dodoi{10.1086/159489}

\bibitem[{{Reeves} \& {Turner}(2000)}]{Reeves&Turner2000}
{Reeves}, J.~N., \& {Turner}, M.~J.~L. 2000, \mnras, 316, 234,
  \dodoi{10.1046/j.1365-8711.2000.03510.x}

\bibitem[{{Ricci} {et~al.}(2017){Ricci}, {Trakhtenbrot}, {Koss}, {Ueda},
  {Delvecchio}, {Treister}, {Schawinski}, {Paltani}, {Oh}, {Lamperti},
  {Berney}, {Gandhi}, {Ichikawa}, {Bauer}, {Ho}, {Asmus}, {Beckmann}, {Soldi},
  {Balokovi{\'c}}, {Gehrels}, \& {Markwardt}}]{Ricci+2017}
{Ricci}, C., {Trakhtenbrot}, B., {Koss}, M.~J., {et~al.} 2017, \apjs, 233, 17,
  \dodoi{10.3847/1538-4365/aa96ad}

\bibitem[{{Risaliti} \& {Lusso}(2015)}]{Risaliti&Lusso2015}
{Risaliti}, G., \& {Lusso}, E. 2015, \apj, 815, 33,
  \dodoi{10.1088/0004-637X/815/1/33}

\bibitem[{{Risaliti} {et~al.}(2009){Risaliti}, {Young}, \&
  {Elvis}}]{Risaliti+2009}
{Risaliti}, G., {Young}, M., \& {Elvis}, M. 2009, \apjl, 700, L6,
  \dodoi{10.1088/0004-637X/700/1/L6}

\bibitem[{{Ross} \& {Fabian}(1993)}]{Ross&Fabian1993}
{Ross}, R.~R., \& {Fabian}, A.~C. 1993, \mnras, 261, 74,
  \dodoi{10.1093/mnras/261.1.74}

\bibitem[{{Ross} \& {Fabian}(2005)}]{Ross&Fabian2005}
{Ross}, R.~R., \& {Fabian}, A.~C. 2005, \mnras, 358, 211,
  \dodoi{10.1111/j.1365-2966.2005.08797.x}

\bibitem[{{R{\'o}{\.z}a{\'n}ska} {et~al.}(2015){R{\'o}{\.z}a{\'n}ska},
  {Malzac}, {Belmont}, {Czerny}, \& {Petrucci}}]{Rozanska2015}
{R{\'o}{\.z}a{\'n}ska}, A., {Malzac}, J., {Belmont}, R., {Czerny}, B., \&
  {Petrucci}, P.~O. 2015, \aap, 580, A77, \dodoi{10.1051/0004-6361/201526288}

\bibitem[{{Shemmer} {et~al.}(2008){Shemmer}, {Brandt}, {Netzer}, {Maiolino}, \&
  {Kaspi}}]{Shemmer+2008}
{Shemmer}, O., {Brandt}, W.~N., {Netzer}, H., {Maiolino}, R., \& {Kaspi}, S.
  2008, \apj, 682, 81, \dodoi{10.1086/588776}

\bibitem[{{Shu} {et~al.}(2010){Shu}, {Yaqoob}, \& {Wang}}]{Shu+2010}
{Shu}, X.~W., {Yaqoob}, T., \& {Wang}, J.~X. 2010, \apjs, 187, 581,
  \dodoi{10.1088/0067-0049/187/2/581}

\bibitem[{{Singh} {et~al.}(1985){Singh}, {Garmire}, \& {Nousek}}]{Singh+1985}
{Singh}, K.~P., {Garmire}, G.~P., \& {Nousek}, J. 1985, \apj, 297, 633,
  \dodoi{10.1086/163560}

\bibitem[{{Sobolewska} {et~al.}(2009){Sobolewska}, {Gierli{\'n}ski}, \&
  {Siemiginowska}}]{Sobolewska+2009a}
{Sobolewska}, M.~A., {Gierli{\'n}ski}, M., \& {Siemiginowska}, A. 2009, \mnras,
  394, 1640, \dodoi{10.1111/j.1365-2966.2009.14436.x}

\bibitem[{{Steffen} {et~al.}(2006){Steffen}, {Strateva}, {Brandt}, {Alexander},
  {Koekemoer}, {Lehmer}, {Schneider}, \& {Vignali}}]{Steffen+2006}
{Steffen}, A.~T., {Strateva}, I., {Brandt}, W.~N., {et~al.} 2006, \aj, 131,
  2826, \dodoi{10.1086/503627}

\bibitem[{{Tange}(2015)}]{Tange2011}
{Tange}, O. 2015, {GNU Parallel 20150322 ('Hellwig')},  Zenodo,
  \dodoi{10.5281/zenodo.16303}

\bibitem[{Taylor(2005)}]{Taylor2005}
Taylor, M.~B. 2005, in Astronomical Society of the Pacific Conference Series,
  Vol. 347, Astronomical Data Analysis Software and Systems XIV, ed.
  P.~Shopbell, M.~Britton, \& R.~Ebert, 29

\bibitem[{{Trakhtenbrot} {et~al.}(2017){Trakhtenbrot}, {Ricci}, {Koss},
  {Schawinski}, {Mushotzky}, {Ueda}, {Veilleux}, {Lamperti}, {Oh}, {Treister},
  {Stern}, {Harrison}, {Balokovi{\'c}}, \& {Gehrels}}]{Trakhtenbrot+2017a}
{Trakhtenbrot}, B., {Ricci}, C., {Koss}, M.~J., {et~al.} 2017, \mnras, 470,
  800, \dodoi{10.1093/mnras/stx1117}

\bibitem[{{Tripathi} {et~al.}(2019){Tripathi}, {Waddell}, {Gallo}, {Welsh}, \&
  {Chiang}}]{Tripathi+2019}
{Tripathi}, S., {Waddell}, S.~G.~H., {Gallo}, L.~C., {Welsh}, W.~F., \&
  {Chiang}, C.~Y. 2019, \mnras, 488, 4831, \dodoi{10.1093/mnras/stz1988}

\bibitem[{{Ursini} {et~al.}(2020){Ursini}, {Petrucci}, {Bianchi}, {Matt},
  {Middei}, {Marcel}, {Ferreira}, {Cappi}, {De Marco}, {De Rosa}, {Malzac},
  {Marinucci}, {Ponti}, \& {Tortosa}}]{Ursini+2020a}
{Ursini}, F., {Petrucci}, P.~O., {Bianchi}, S., {et~al.} 2020, \aap, 634, A92,
  \dodoi{10.1051/0004-6361/201936486}

\bibitem[{{Vasudevan} \& {Fabian}(2007)}]{Vasudevan&Fabian2007}
{Vasudevan}, R.~V., \& {Fabian}, A.~C. 2007, \mnras, 381, 1235,
  \dodoi{10.1111/j.1365-2966.2007.12328.x}

\bibitem[{{Vasudevan} \& {Fabian}(2009)}]{Vasudevan&Fabian2009}
{Vasudevan}, R.~V., \& {Fabian}, A.~C. 2009, \mnras, 392, 1124,
  \dodoi{10.1111/j.1365-2966.2008.14108.x}

\bibitem[{{Waddell} \& {Gallo}(2020)}]{Waddell&Gallo2020}
{Waddell}, S.~G.~H., \& {Gallo}, L.~C. 2020, \mnras, 498, 5207,
  \dodoi{10.1093/mnras/staa2783}

\bibitem[{{Waddell} {et~al.}(2024){Waddell}, {Nandra}, {Buchner}, {Wu}, {Shen},
  {Arcodia}, {Merloni}, {Salvato}, {Dauser}, {Boller}, {Liu}, {Comparat},
  {Wolf}, {Dwelly}, {Ricci}, {Brownstein}, \& {Brusa}}]{Waddell+2023}
{Waddell}, S.~G.~H., {Nandra}, K., {Buchner}, J., {et~al.} 2024, \aap, 690,
  A132, \dodoi{10.1051/0004-6361/202245572}

\bibitem[{Walter \& Fink(1993)}]{Walter&Fink1993}
Walter, R., \& Fink, H.~H. 1993, Astronomy \& Astrophysics, 274, 105

\bibitem[{{Wilkins} {et~al.}(2021){Wilkins}, {Gallo}, {Costantini}, {Brandt},
  \& {Blandford}}]{Wilkins+2021}
{Wilkins}, D.~R., {Gallo}, L.~C., {Costantini}, E., {Brandt}, W.~N., \&
  {Blandford}, R.~D. 2021, \nat, 595, 657, \dodoi{10.1038/s41586-021-03667-0}

\bibitem[{{Xiang} {et~al.}(2022){Xiang}, {Ballantyne}, {Bianchi}, {De Rosa},
  {Matt}, {Middei}, {Petrucci}, {R{\'o}{\.z}a{\'n}ska}, \&
  {Ursini}}]{Xiang+2022}
{Xiang}, X., {Ballantyne}, D.~R., {Bianchi}, S., {et~al.} 2022, \mnras, 515,
  353, \dodoi{10.1093/mnras/stac1646}

\bibitem[{{Xu} {et~al.}(2021{\natexlab{a}}){Xu}, {Ding}, {Gu}, {Guo}, \&
  {Contini}}]{Xu+2021}
{Xu}, X., {Ding}, N., {Gu}, Q., {Guo}, X., \& {Contini}, E. 2021{\natexlab{a}},
  \mnras, 507, 3572, \dodoi{10.1093/mnras/stab2278}

\bibitem[{{Xu} {et~al.}(2021{\natexlab{b}}){Xu}, {Garc{\'\i}a}, {Walton},
  {Connors}, {Madsen}, \& {Harrison}}]{Xu+2021a}
{Xu}, Y., {Garc{\'\i}a}, J.~A., {Walton}, D.~J., {et~al.} 2021{\natexlab{b}},
  \apj, 913, 13, \dodoi{10.3847/1538-4357/abf430}

\bibitem[{{Young} {et~al.}(2010){Young}, {Elvis}, \& {Risaliti}}]{Young+2010}
{Young}, M., {Elvis}, M., \& {Risaliti}, G. 2010, \apj, 708, 1388,
  \dodoi{10.1088/0004-637X/708/2/1388}

\bibitem[{{Yuan} \& {Narayan}(2014)}]{Yuan&Narayan2014}
{Yuan}, F., \& {Narayan}, R. 2014, \araa, 52, 529,
  \dodoi{10.1146/annurev-astro-082812-141003}

\bibitem[{{Zhang} {et~al.}(2025){Zhang}, {Santangelo}, {Xu}, {Feng}, {Lu},
  {Chen}, {Ge}, {Nandra}, {Wu}, {Feroci}, {Hernanz}, {Liu}, {He}, {Wang},
  {Jiang}, {Cui}, {Yang}, {Wang}, {Li}, {Liu}, {Meng}, {Wen}, {Zhang}, {Ma},
  {Li}, {Li}, {Qi}, {Sun}, {Luo}, {Liu}, {Liu}, {Zhang}, {Luo}, {Zhu}, {Zhao},
  {Sun}, {Yang}, {Wu}, {Jiang}, {Shi}, {Liu}, {Xu}, {Yang}, {Zhang}, {Han},
  {Gao}, {Huo}, {Zhang}, {Wang}, {Zhao}, {Cui}, {Wang}, {Wang}, {Li}, {Bao},
  {Liu}, {Wang}, {Wang}, {Wang}, {Wang}, {Wang}, {Ding}, {Sheng}, {Qiang},
  {Yan}, {Liu}, {Wu}, {Liu}, {Chen}, {Zhang}, {Liu}, {Altmann}, {Bechteler},
  {Burwitz}, {Fiorini}, {Friedrich}, {Meidinger}, {Strecker}, {Baldini},
  {Bellazzini}, {Bonino}, {Frass{\`a}}, {Latronico}, {Maldera}, {Manfreda},
  {Minuti}, {Pesce-Rollins}, {Sgr{\`o}}, {Tugliani}, {Pareschi}, {Basso},
  {Sironi}, {Spiga}, {Tagliaferri}, {Tykhonov}, {Paltani}, {Bozzo}, {Tenzer},
  {Bayer}, {Tuo}, {Liu}, {Zhang}, {Cai}, {Liu}, {Chen}, {Wang}, {He}, {Chen},
  {Qiu}, {Zhang}, {Feng}, {Zhu}, {Zhou}, {Zheng}, {Song}, {Shi}, {Wang}, {Jia},
  {Jiang}, {Li}, {Zhao}, {Guan}, {Zhang}, {Li}, {Huang}, {Liao}, {You},
  {Zhang}, {Wang}, {Wang}, {Ou}, {Hu}, {Shi}, {Cui}, {Jiang}, {Cheng}, {Li},
  {Xu}, {Zane}, {Bambi}, {Bu}, {Dall'Osso}, {De Rosa}, {Gou}, {Guillot}, {Ji},
  {Li}, {Mao}, {Patruno}, {Stratta}, {Taverna}, {Tsygankov}, {Uttley}, {Watts},
  {Wu}, {Xu}, {Yi}, {Zhang}, {Zhang}, {Zhao}, \& {Zhou}}]{Zhang+2025a}
{Zhang}, S.-N., {Santangelo}, A., {Xu}, Y., {et~al.} 2025, arXiv e-prints,
  arXiv:2506.08101, \dodoi{10.48550/arXiv.2506.08101}

\bibitem[{{Zhou} {et~al.}(2025){Zhou}, {Mao}, {Zhang}, {Patruno}, {Bozzo},
  {Xu}, {Santangelo}, {Zane}, {Zhang}, {Feng}, {Cavecchi}, {De Marco}, {Fan},
  {Hou}, {Jiang}, {Romano}, {Sala}, {Tao}, {Veledina}, {Vink}, {Wang}, {Wang},
  {Wang}, {Weng}, {Wu}, {Xie}, {Zhang}, {Zhang}, {Zhao}, {Zheng}, {Barua},
  {Chen}, {Chen}, {Chen}, {Chen}, {Chen}, {Cheng}, {Chi}, {Cui}, {de Martino},
  {Deng}, {Ducci}, {Farinelli}, {Feng}, {Ge}, {Gu}, {Guo}, {Han}, {Hu},
  {Huang}, {in't Zand}, {Ji}, {Kang}, {Kini}, {Li}, {Li}, {Liu}, {Liu}, {Liu},
  {Lyu}, {Marino}, {Markowitz}, {Mezcua}, {Middleton}, {Mou}, {Ng}, {Papitto},
  {Pei}, {Peng}, {Poutanen}, {Shui}, {Simone}, {Su}, {Tan}, {Wang}, {Wang},
  {Wang}, {Wang}, {Wang}, {Wang}, {Wang}, {Wu}, {Xiao}, {Xiong}, {Xu}, {Xue},
  {Yan}, {Yang}, {Yang}, {Yang}, {Ye}, {Yu}, {Yuan}, {Zhang}, {Zhang}, {Zhao},
  {Zhao}, {Zheng}, {Zheng}, \& {Zuo}}]{Zhou+2025}
{Zhou}, P., {Mao}, J., {Zhang}, L., {et~al.} 2025, arXiv e-prints,
  arXiv:2506.08367, \dodoi{10.48550/arXiv.2506.08367}

\bibitem[{{Zoghbi} {et~al.}(2015){Zoghbi}, {Miller}, {Walton}, {Harrison},
  {Fabian}, {Reynolds}, {Boggs}, {Christensen}, {Craig}, {Hailey}, {Stern}, \&
  {Zhang}}]{Zoghbi+2015}
{Zoghbi}, A., {Miller}, J.~M., {Walton}, D.~J., {et~al.} 2015, \apjl, 799, L24,
  \dodoi{10.1088/2041-8205/799/2/L24}

\end{thebibliography}
\bibliographystyle{aasjournal}

\end{document}